\documentclass{iopart}
\usepackage{iopams, setstack}
\usepackage{bm,hyperref}
\usepackage{graphicx,wrapfig}

\newtheorem{theorem}{Theorem}
\newtheorem{corollary}{Corollary}
\newtheorem{definition}{Definition}

\newcommand{\eqref}[1]{{(\ref{#1})}}    

\newcommand{\ItP}{{\mathcal{P}}} 
\newcommand{\Lie}{{\mathcal{L}}}  
\newcommand{\LieS}{{\mathcal{L}_{\psi}}} 
\newcommand{\LieX}{{\mathcal{L}_{\xi}}} 

\newcommand{\sfa}{{\mathsf{a}}}  
\newcommand{\sfb}{{\mathsf{b}}}
\newcommand{\sfc}{{\mathsf{c}}}
\newcommand{\sfd}{{\mathsf{d}}}
\newcommand{\sff}{{\mathsf{f}}}

\begin{document}
\title[Approximate spacetime symmetries]{Approximate spacetime symmetries and conservation laws}
\author{Abraham I. Harte}
\address{Enrico Fermi Institute}
\address{University of Chicago, Chicago, IL, 60637, USA}
\ead{harte@uchicago.edu}

\date{July 31, 2008}

\begin{abstract}
A notion of geometric symmetry is introduced that generalizes the
classical concepts of Killing fields and other affine collineations.
There is a sense in which flows under these new vector fields
minimize deformations of the connection near a specified observer.
Any exact affine collineations that may exist are special cases. The
remaining vector fields can all be interpreted as analogs of
Poincar\'{e} and other well-known symmetries near timelike
worldlines. Approximate conservation laws generated by these objects
are discussed for both geodesics and extended matter distributions.
One example is a generalized Komar integral that may be taken to
define the linear and angular momenta of a spacetime volume as seen
by a particular observer. This is evaluated explicitly for a
gravitational plane wave spacetime.
\end{abstract}


\vskip 1pc

\section{Introduction}

It is rarely possible to model realistic physical systems with exact
solutions to the equations of some general underlying theory.
Despite this, many interesting problems deviate only slightly from a
model problem that can be understood exactly. Such solutions are
usually tractable only because of symmetry assumptions. Once they're
understood, perturbation theory may be used to understand many
systems that ``almost'' satisfy the appropriate symmetry principle.
While this statement has a clear intuitive meaning, quantifying it
can be difficult. It is also not necessarily obvious how -- or if it
is meaningful -- to uniquely propagate a symmetry from solutions
where it is exact into the perturbations that break it. These issues
are particularly important in the context of conservation laws. As
an example, one might want to know how to construct approximately
conserved quantities that are unique generalizations of some exact
counterpart in a similar system. There would hopefully be a sense in
which such quantities varied slowly for some class of small
perturbations.

Some steps towards understanding problems like these are explored
here in the context of affine collineations (of which Killing
vectors are special cases) associated with curved spacetimes. While
these kinds of symmetries rarely exist, there are various senses in
which approximate replacements can usually be introduced. One method
for finding vector fields that are ``almost Killing'' is to write
down an action whose value provides some sense for how nearly a
particular flow preserves the metric \cite{Matzner}. There are
important caveats to this interpretation, although the final result
is that any vector field extremizing such an action satisfies a
fairly simple generalization of Killing's equation. Various reasons
have been given for suggesting other extensions as well
\cite{AlmostStat,YanoBochner,Komar1,Komar2,YorkSym}. While any
genuine Killing vectors that might exist are solutions to all of
these equations, it is not usually clear how the remaining fields
should be interpreted. This problem arises even in flat spacetime.

What form an approximate symmetry should take is highly dependent on
its intended use. One application is in the estimation of a black
hole's angular momentum. This requires finding rotational Killing
fields on certain 2-spheres foliating a horizon. Various methods
have therefore been developed for defining such objects using only
the intrinsic geometry of these surfaces
\cite{S2Conformal,S2KT,S2Killing}. The concept of approximate
Killing fields has also been adapted for use on initial data sets
used in $3+1$ splits of Einstein's equation \cite{DainInitData}.

The approach taken here is to define a set of vector fields in a
four dimensional volume that can all be viewed as analogs of known
symmetries in Minkowski spacetime. The physical interpretation is
that these fields may be viewed as generators of approximate
symmetries by a specified observer. Any sufficiently small region
near a particular point can be made to look nearly flat. Some
structures from Minkowski spacetime may be therefore be introduced
very near this point. Approximate symmetries that take advantage of
this fact are proposed in Sect. \ref{Sect:Symmetries}. It is then
shown in Sect. \ref{Sect:GAC} that analogous objects can also be
introduced near an observer's worldline. These sorts of vector
fields can actually be extended in a non-perturbative way to finite
regions around the point or worldline from which they were
constructed. A well-defined subset provides a precise analog of the
Poincar\'{e} group. Translations, rotations, and boosts very near an
observer extend in a useful way to cover large portions of the
spacetime. Any exact symmetries that may exist are included as
special cases. Some connections to conservation laws are discussed
in Sect. \ref{Sect:ConsLaws}, and a simple example involving
gravitational plane wave is finally presented in Sect.
\ref{Sect:Example}.

\subsection*{Exact symmetries}

Given some spacetime $(\mathcal{M},g_{ab})$, there are several types
of exact symmetries that may be discussed. The most common of these
take the form of vector fields whose associated diffeomorphisms
preserve some geometric structure. The most ubiquitous examples are
the Killing fields. Their flows preserve the metric. Vector fields
$Y^a_{\mathrm{K}}$ with this property satisfy
\begin{equation}
  \Lie_{Y_{\mathrm{K}}} g_{ab} = 0 . \label{KillingDef}
\end{equation}
Any solutions that may exist can be used to find conserved
quantities associated with geodesics or matter distributions
\cite{Wald}, identify mass centers \cite{SchattStreub1}, simplify
Einstein's equation \cite{SimplifyEinstWSyms, ExactSolns}, classify
its solutions \cite{ExactSolns, Hall}, and so on.

This utility has (among other reasons) motivated various
generalizations of \eqref{KillingDef}. Perhaps the simplest of these
arises from considering flows that preserve the metric only up to
some constant factor:
\begin{equation}
  \Lie_{Y_{\mathrm{H}} } g_{ab} = 2 c g_{ab} .
  \label{HomotheticDef}
\end{equation}
Any $Y^a_{\mathrm{H}}$ satisfying this equation with constant $c$ is
known as a homothetic vector field or homothety. Allowing the
dilation factor $c$ to vary would define a conformal Killing vector.
These objects preserve the metric up to an arbitrary multiplicative
factor. The standard Killing vectors are special cases of either
class. Like them, conformal and homothetic vector fields usually do
not exist. Their presence can be very useful, however. The existence
of a proper (non-Killing) homothetic vector is often used to define
a notion of geometric self-similarity, for example. Such objects
therefore appear in certain models of gravitational collapse and
cosmology. They are also related to the appearance of critical
phenomena in general relativity \cite{SelfSimilar}.

A simple generalization of the homotheties can be found by
considering vector fields that satisfy
\begin{equation}
  \nabla_a \Lie_{Y_{\mathrm{A}}} g_{bc} = 0 .
  \label{AffineDef}
\end{equation}
Solutions to this equation are known as affine collineations. They
are the generators of infinitesimal affine transformations. Killing
and homothetic vector fields are special cases. All of the affine
collineations may be interpreted geometrically as preserving the
Levi-Civita connection. This means that $\Lie_{Y_{\mathrm{A}}}$ and
$\nabla_a$ commute when acting on arbitrary tensor fields. Geodesics
and their affine parameters are also preserved under the action of
any $Y_{\mathrm{A}}^a$. Although this might seem to be a significant
generalization of the Killing vector concept, solutions rarely
exist. The only non-flat vacuum spacetimes that admit non-homothetic
affine collineations are the \textit{pp}-waves \cite{RareAffine}.
Similarly, it has been shown that proper homotheties cannot exist in
any asymptotically flat vacuum spacetime with positive Bondi mass
\cite{RareHomothety}. Despite these results, interesting affine
collineations can occasionally be identified in geometries that are
not Ricci-flat. Doing so provides a number of simplifications for
various problems. Some of these derive from the fact that
\begin{equation}
  K_{ab} = \Lie_{Y_{\mathrm{A}}} g_{ab} \label{AffineKT}
\end{equation}
is a second-rank Killing tensor; i.e.
\begin{equation}
  \nabla_{(a} K_{bc)} = 0. \label{KTDef}
\end{equation}
It should be noted that not all symmetric tensors satisfying this
equation can be derived from affine collineations. One
counterexample is the Killing tensor associated with Carter
constants in Kerr.

Transformations generated by affine collineations can be viewed as
mapping geodesics into geodesics. They preserve the affine
parameters of each curve. Dropping this latter requirement recovers
the so-called projective collineations. A precise definition may be
found in \cite{Hall}, although it will not be needed here. One of
their interesting consequences is that they leave invariant the
projective curvature tensor:
\begin{equation}
  \Lie_{Y_{\mathrm{P}}} \big( R^{a}{}_{bcd} - \frac{2}{3} \delta^{a}_{[c} R_{d]b} \big) = 0.
\end{equation}
Vector fields satisfying this equation are not always projective,
however.

The list of definitions here could keep growing as new fields are
added that preserve more and more geometric structures.
Interestingly, the quantities introduced so far all share a very
useful characteristic that does not easily generalize: the space of
vector fields in each of the mentioned classes has finite dimension.
Furthermore, any single element is uniquely determined by its value
and the values of its first one or two derivatives at a single
point. These properties are well-known for Killing fields. Four
dimensional spacetimes (which is all that will be considered here)
admit a maximum of 10 linearly independent Killing vectors. At most
one homothety can exist that is not itself Killing. The maximum
number of (not necessarily proper) conformal Killing vectors is 15,
and the affine collineations total no more than 20. Finally, the
space of projective collineations has a maximum of 24 dimensions
\cite{Hall}. Properties like these do not hold for vector fields
whose flows leave invariant the Riemann, Ricci, or Einstein
curvature tensors of a given spacetime. Despite this, the class of
approximate symmetries introduced below is constructed so as to have
finite dimension. Any given member is fixed by its value together
with the values of its first derivatives at a point. Unlike the
exact symmetries, these objects always exist at least in some finite
region. After fixing a reference frame, the space of approximate
symmetries has exactly 20 dimensions. Ten of these will be
identifiable as generalized Killing vectors, while the remaining ten
will be related to more general affine collineations.

It has already been remarked that the presence of Killing fields
implies the existence of various conserved quantities. The same can
also be said for more general collineations. Extensive discussions
of exact symmetries and associated integrals of the geodesic
equation may be found in \cite{GeodesicConsts}. Many of these
symmetries are non-Noetherian in the sense that they preserve the
equations of motion, but not the action. Despite this, their
presence allows constants of motion to also be assigned to arbitrary
stress-energy distributions satisfying Einstein's equation
\cite{KatzinLevine, Collinson}. As will be discussed in Sect.
\ref{Sect:ConsLaws}, generalizations of these quantities can be
associated with any approximate affine collineations that are
identified.

\section{Symmetries near a point}\label{Sect:Symmetries}

Generic symmetries in general relativity are usually discussed in
the context of asymptotically flat spacetimes. There then exist
approximate notions of isometry that improve as one approaches
infinity \cite{Wald}. Generalizations of these ideas also exist for
geometries with somewhat more complicated (but still highly
symmetric) asymptotic behavior like that of anti-de Sitter
\cite{AsymptAdS}. The assumption of a simple limiting form for the
geometry makes it convenient to invariantly describe certain
properties of a spacetime in terms of ``measurements at infinity.''
Quantities that may be identified as a spacetime's total energy or
angular momentum appear naturally, for example.

While useful in many contexts, these ideas do not always translate
into observations made by physical observers. Measurements like
those expected from gravitational wave detectors do come very close
to fitting into this formalism. Others can require a more local
description. In particular, it is sometimes important to understand
what given observers would experience inside strongly curved regions
of spacetime. Abstracting the concept of an observer to a timelike
worldline $\Gamma$, vector fields may be introduced in (say) some
convex neighborhood $W$ of $\Gamma$ that act like approximations to
Killing fields or more general collineations. This is always
possible, and the symmetries these vectors generalize become exact
on $\Gamma$ itself. Limiting collineations can evidently be useful
on scales that are either very large or very small. It is much less
clear how to easily describe systems at intermediate distances.

\subsection{Motivation}

The idea of a local symmetry just outlined is best introduced by
first considering vector fields $\psi^a(x,\gamma)$ that generalize
the affine collineations in some reasonable way near a fixed
reference point $\gamma$. Let these objects be defined inside a
normal neighborhood $N$ of this point. It is intuitively obvious
that vector fields may always be chosen such that $\nabla_a \LieS
g_{bc}$ vanishes at $\gamma$. While this condition is reasonable to
require, it is not very interesting by itself. Much more can be said
if each vector field in this class is uniquely fixed in $N$ by
knowledge of
\begin{equation}
  \psi^\sfa(\gamma,\gamma) , \qquad \nabla_
  \sfa \psi^\sfb(\gamma,\gamma) . \label{InitDataFirst}
\end{equation}
It will be assumed that each $\psi^a$ depends linearly on this
initial data with no degeneracy. This implies that there are always
$4+16=20$ linearly independent vector fields defined about any given
point in a four dimensional spacetime. Note that indices in
\eqref{InitDataFirst} have been written in a sans-serif font to
emphasize that they are associated with the preferred point
$\gamma$.

Approximate affine collineations with the appropriate properties may
be constructed by projecting symmetries of the tangent space
$T_\gamma N$ into $N$ using the exponential map. Consider the linear
transformations
\begin{equation}
  X^\sfa \rightarrow X^\sfa + \epsilon B_{\sfb}{}^{a} X^\sfb
  \label{XFormVect}
\end{equation}
of vectors $X^\sfa$ in this space parameterized by an arbitrary
tensor $B_{\sfb}{}^{\sfa}$. Being a vector space, $T_\gamma N$ has a
preferred origin. Adding another term to \eqref{XFormVect} to shift
that origin would be awkward. The translational symmetries that such
a procedure might produce are certainly important, although
generating them requires a more subtle treatment described below.
For now, consider only the vector fields
\begin{equation}
  \Psi^\sfa = X^\sfb B_{\sfb}{}^{\sfa}
  \label{AffineTangent}
\end{equation}
associated with homogeneous transformations of the given form. These
clearly satisfy
\begin{equation}
  \frac{\partial}{\partial X^\sfa} \Lie_\Psi g_{\sfb \sfc} = 2 \frac{\partial}{\partial
  X^\sfa} \left( g_{\sfd(\sfc}(\gamma) \frac{\partial}{\partial X^{\sfb)}} \Psi^{\sfd} \right)
  =0,
\end{equation}
so they are affine within $T_\gamma N$ in the sense of
\eqref{AffineDef}. Such transformations can be made to induce shifts
$x \rightarrow x + \epsilon \psi$ in spacetime points associated
with vectors $X^\sfa$ via
\begin{equation}
  x = \exp_\gamma X .
  \label{ExpMap}
\end{equation}

A simple relation between $\psi^a$ and $\Psi^\sfa$ is found by
introducing Synge's world function $\sigma(x,y) = \sigma(y,x)$. This
two-point scalar returns one-half of the squared geodesic distance
between its arguments. The assumption that $N$ be a normal
neighborhood of $\gamma$ ensures that $\sigma(x,\gamma)$ is uniquely
defined for all points $x$ in this region. Many of its properties
are reviewed in \cite{Synge, PoissonRev}. Most importantly for the
problem at hand, the first derivative of the world function
effectively inverts the exponential map. Any set $\{ \gamma, x, X^a
\}$ satisfying \eqref{ExpMap} is related via
\begin{equation}
  X_{\sfa} = - \sigma_{\sfa}(x,\gamma),
  \label{XDef}
\end{equation}
where the common shorthand $\sigma_\sfa = \nabla_\sfa \sigma =
\partial \sigma/ \partial \gamma^\sfa$ has been used. The
right-hand side of (\ref{XDef}) generalizes the concept of a
separation vector between two points. It is useful in that a
straightforward expansion shows that linear transformations of the
form (\ref{XFormVect}) effectively shift spacetime points by an
amount parameterized with a vector $\psi^{a}(x,\gamma)$ satisfying
\begin{equation}
\Psi^{\sfa} = - \sigma^{\sfa}{}_{a} \psi^{a} . \label{PsiTopsi}
\end{equation}
If the various components of $X^\sfa$ as defined in (\ref{XDef}) are
used as coordinates, the bitensor $-\sigma^{\sfa}{}_{a} = - g^{\sfa
\sfb}
\partial^2 \sigma/ \partial x^a \partial \gamma^{\sfb}$
reduces to the identity. Components of $\psi^a$ and $\Psi^{\sfa}$
are therefore identical in normal coordinate systems of this type.
In general, it is useful to introduce
\begin{equation}
  H^{a}{}_{\sfa} = [ - \sigma^{\sfa}{}_{a} ]^{-1} \label{HDef}
\end{equation}
as the matrix inverse of the operator appearing in \eqref{PsiTopsi}.
This always exists in the regions considered here. Using
\eqref{AffineTangent} now shows that
\begin{equation}
 \psi^{a}(x,\gamma) = - H^{a}{}_{\sfa} \sigma_{\sfb} B^{\sfb \sfa} .
 \label{PsiPart}
\end{equation}
Holding $\gamma$ fixed, this equation defines a 16-parameter family
of vector fields generated by $B_{\sfa \sfb}=\nabla_\sfa \psi_\sfb
(\gamma,\gamma)$. Every such $\psi^a$ vanishes at $\gamma$. It also
satisfies \eqref{AffineDef} at this point. In flat spacetime, these
vector fields coincide everywhere with exact affine collineations.

Not all such symmetries are included in (\ref{PsiPart}), however.
The four translational Killing fields are missing. These can be
obtained by considering transformations that directly shift the base
point $\gamma$. Perturbations of this form cannot leave $X^\sfa$
fixed, as the initial and final vectors must be elements of
different spaces. Introducing some $A^\sfa$, we therefore demand
that $X^\sfa$ be parallel-transported along the curve that $\gamma$
follows under the one-parameter family of transformations
\begin{equation}
  \gamma \rightarrow \gamma + \epsilon A .
\end{equation}
Using this together with \eqref{XDef} and the homogeneous
transformation \eqref{XFormVect} generates the full 20-parameter
family of approximate affine collineations
\begin{equation}
  \psi^{a} = H^{a}{}_{\sfa} ( \sigma^{\sfa}{}_{\sfb} A^\sfb - \sigma_\sfb B^{\sfb \sfa}
  ) . \label{JacobiFirstDer}
\end{equation}
Given any $A^{\sfa}$ and $B^{\sfa \sfb}$, these objects all satisfy
\begin{equation}
  \nabla_\sfa \LieS g_{\sfb \sfc} (\gamma) = 0 . \label{JacobiAffine}
\end{equation}
The initial data
\begin{equation}
  A^\sfa = \psi^\sfa(\gamma,\gamma) , \qquad B^{\sfa \sfb} =
  \nabla^\sfa
  \psi^\sfb(\gamma,\gamma) \label{InitialData}
\end{equation}
determine $\psi^a(x,\gamma)$ throughout $N$. In Minkowski spacetime,
one finds that
\begin{equation}
    \psi^\alpha = A^\alpha + (x-\gamma)^\beta B_{\beta}{}^{\alpha}
\end{equation}
in the usual coordinates. These coincide exactly with all of the
affine collineations in this geometry.

In general, vector fields satisfying (\ref{AffineDef}) in a curved
spacetime also have the form (\ref{JacobiFirstDer}) for some
$A^\sfa$ and $B^{\sfa \sfb}$. This is most easily seen by noting
that vector fields with the given form have been obtained before as
general solutions to the equation of geodesic deviation (also known
as the Jacobi equation) \cite{Dix70a}
\begin{equation}
  \sigma^{b} \sigma^{c} ( \nabla_{b} \nabla_{c} \psi_{a} -
  R_{abc}{}^{d} \psi_{d} ) = 0. \label{Jacobi}
\end{equation}
For any fixed $x$, this is an ordinary differential equation along
the geodesic connecting that point to $\gamma$. Solving it
repeatedly for all geodesics in $N$ passing through this origin
reproduces the vector fields (\ref{JacobiFirstDer}). It is clear
that such solutions always exist as long as the geometry is
reasonably smooth. These are the spacetime's Jacobi fields about
$\gamma$. The bitensors $H^{a}{}_{\sfa} \sigma^{\sfa}{}_{\sfb}$ and
$H^{a}{}_{\sfa} \sigma_\sfb$ are known as Jacobi propagators.

Solutions to the geodesic deviation equation effectively map one
geodesic into another while preserving the affine parameters of both
curves. It was noted above that this is the defining characteristic
of affine collineations. The difference is that such vector fields
must map \textit{every} geodesic into another geodesic. This
intuitive argument makes it clear that all affine collineations --
or Killing fields as special cases -- must be solutions of
(\ref{Jacobi}). The proof follows from noting that second
derivatives of any exact affine collineation $Y_{\mathrm{A}}^a(x)$
must satisfy
\begin{equation}
  \nabla_{b} \nabla_{c} Y^a_{\mathrm{A}} = - R_{bdc}{}^{a} Y^d_{\mathrm{A}} .
    \label{Del2Affine}
\end{equation}
This result is clear from \eqref{Del2General}, and is actually
equivalent to \eqref{AffineDef}. Substituting it into (\ref{Jacobi})
shows that all affine collineations are indeed special cases of
Jacobi fields. As expected, $Y^a_{\mathrm{A}}$ satisfies the
geodesic deviation equation along all geodesics; even those that do
not pass through $\gamma$. This point illustrates precisely how
\eqref{Jacobi} generalizes the equation defining an affine
collineation. It is simply \eqref{Del2Affine} contracted into
$\sigma^b \sigma^c$. Alternatively, the Jacobi equation is
equivalent to \eqref{LieSig7}.

To summarize, the following is now evident:
\begin{theorem}\label{Thm:JacobiBasic}
  Let $N$ be a normal neighborhood of some point $\gamma$. Define a Jacobi
  field $\psi^a(x,\gamma)$ to be a solution of \eqref{Jacobi} throughout this region. It is explicitly given by
  \eqref{JacobiFirstDer} for some initial data with the form
  \eqref{InitialData}. The set of all Jacobi fields about a fixed $\gamma$ forms a 20-dimensional group in
  four spacetime dimensions. Each element satisfies $\mathcal{L}_\psi \nabla =0$ at
  $\gamma$. Furthermore, all affine collineations are members of
  this group.
\end{theorem}
These properties motivate our identification of the Jacobi fields as
generalizations of affine collineations near $\gamma$.

Further results that strengthen this decision are derived in the
appendix. Even though few Jacobi fields are genuine affine
collineations, all can be interpreted as exact symmetries of certain
quantities connected with the spacetime's geometric structure:
\begin{theorem}\label{Thm:JacobiExactSyms}
  Given a Jacobi field defined as in theorem \ref{Thm:JacobiBasic},
  it is always true that
  \begin{equation}
    \LieS \sigma^\sfa = \LieS \sigma^a = \LieS \sigma^{\sfa}{}_{a} =
    \LieS H^{a}{}_\sfa =0 ,
  \end{equation}
  where one of the arguments in each of these equations is taken to be the
  origin $\gamma$.
\end{theorem}
Lie derivatives on two-point tensor fields are defined to act
independently on each argument. See \eqref{LieSig1}, for example.
Quantities appearing in this theorem are all important in Riemann
normal coordinate systems parameterizing arbitrary points $x$ by the
components of $X^\sfa = -\sigma^\sfa (x,\gamma)$. In terms of a more
direct interpretation of the Jacobi fields as approximately
satisfying \eqref{AffineDef}, Lie derivatives of the metric with
respect to an arbitrary $\psi^a$ are strongly constrained by the
identities \eqref{LieSig6}-\eqref{LieSig9}.

Statements of this sort do not exhaust the connections between the
Jacobi equation and a spacetime's symmetries. There is a sense in
which higher-rank Killing tensors that may exist are also solutions
to the geodesic deviation equation \cite{CavigliaBasic}.
Furthermore, projective collineations can be shown solve an
inhomogeneous form of (\ref{Jacobi}) proportional to $\sigma_a$
\cite{Caviglia}. These observations will not be discussed any
further here, although it is possible that they could be used to
generalize the present framework.

\subsection{Special cases} \label{Sect:SpecialJacobi}

It is often useful to single out a subset of the Jacobi fields
distinguished by antisymmetric $B_{\sfa \sfb} = \nabla_\sfa
\psi_\sfb$. These may be said to generalize only the Killing fields
of a given spacetime. Distinguishing them with a subscript ``K,''
they clearly satisfy $\Lie_{\psi_{\mathrm{K}}} g_{\sfa \sfb}(\gamma)
= 0$ as well as (\ref{JacobiAffine}). Such objects form a
10-dimensional group that may be thought of as a generalization of
the Poincar\'{e} group. They have been suggested before as useful
generators for the linear and angular momenta of extended matter
distributions \cite{Dix70a, Dix74, Dix79}. Fixing a hypersurface
$\Sigma$ that passes through $\gamma$ and the worldtube of some
well-behaved spatially-compact stress-energy distribution $T^{ab}$,
let
\begin{equation}
  p_\sfa(\gamma,\Sigma) A^\sfa + \frac{1}{2} S_{\sfa\sfb}(\gamma,\Sigma) B^{[\sfa\sfb]} = \int_\Sigma T^{a}{}_{b}
  \psi^b_{\mathrm{K}} \rmd S_a .
  \label{DixMomenta}
\end{equation}
Varying the 10 free parameters here determines the four linear
momenta $p^\sfa$ and six angular momenta $S^{\sfa\sfb} =
S^{[\sfa\sfb]}$. Explicit formulae are easily found using
(\ref{JacobiFirstDer}). They coincide with standard definitions in
flat spacetime (where all $\psi^a_{\mathrm{K}}$ are Killing).

Theorem \ref{Thm:JacobiExactSyms} is easily expanded for these
vector fields:
\begin{corollary}\label{Thm:JacobiExactSymsKilling}
    Given any Jacobi field $\psi^a_{\mathrm{K}}$ satisfying
     $\mathcal{L}_{\psi_{\mathrm{K}}} g_{\sfa \sfb} =0$,
    \begin{equation}
      \mathcal{L}_{\psi_{\mathrm{K}}} \sigma =
      \mathcal{L}_{\psi_{\mathrm{K}}} \sigma_\sfa =
      \mathcal{L}_{\psi_{\mathrm{K}}} \sigma_a =
      \mathcal{L}_{\psi_{\mathrm{K}}} \sigma_{\sfa a} =
      \mathcal{L}_{\psi_{\mathrm{K}}} H^{a \sfa} = 0.
    \end{equation}
    Again, one argument in each of these equations is assumed to be $\gamma$.
\end{corollary}
This follows from the well-known identity \cite{Synge, PoissonRev}
\begin{equation}
  \sigma^\sfa \sigma_\sfa = \sigma^{a} \sigma_{a} = 2 \sigma ,
  \label{SigIdent2}
\end{equation}
and its first derivative
\begin{equation}
  \sigma^{\sfa}{}_{a} \sigma^{a} = \sigma^\sfa .
  \label{SigIdent1}
\end{equation}
By definition, $\sigma = g_{\sfa \sfb}(\gamma) X^\sfa X^\sfb/2$ is
one-half of the squared geodesic distance between $\gamma$ and $x$.
Killing-type Jacobi fields based at $\gamma$ therefore drag both
arguments of $\sigma(x,\gamma)$ in such a way that distances are
preserved.

It is also possible to identify Jacobi fields that act like
homotheties near $\gamma$. These are distinguished by letting
\begin{equation}
  B_{(\sfa \sfb)} = \frac{1}{2} \Lie_{\psi_{\mathrm{H}}} g_{\sfa \sfb}(\gamma) = c g_{\sfa
  \sfb} (\gamma). \label{HomB}
\end{equation}
As in \eqref{HomotheticDef}, $c$ is an arbitrary constant. For
simplicity, the purely Killing components of some prospective
$\psi^a_{\mathrm{H}}$ can be removed by setting $A_\sfa = B_{[\sfa
\sfb]} = 0$ and $c \neq 0$. Substitution into (\ref{JacobiFirstDer})
then shows that
\begin{equation}
  \psi^{a}_{\mathrm{H}} = - c H^{a}{}_{\sfa} \sigma^\sfa  = c
  \sigma^{a}. \label{JacobiHom}
\end{equation}
This second equality follows from contracting $\delta^{a}_{b} =
-H^{a}{}_{\sfa} \sigma^{\sfa}{}_{b}$ with $\sigma^{b}$ and using
\eqref{SigIdent1}.

The simplicity of (\ref{JacobiHom}) is interesting, although perhaps
not surprising. It is consistent with the interpretation of
$-\sigma^\sfa$ as a ``separation vector'' between $x$ and $\gamma$.
As has been noted before, $\sigma^{\sfa}{}_{b} \rightarrow -
\delta^{\alpha}_{\beta}$ in a normal coordinate system. The
components of $\sigma^{a}$ would therefore be equal to $X^\alpha$.
The normal coordinate functions themselves act as components of an
approximately homothetic vector field. This is unique up to a
constant factor and the addition of Killing-type Jacobi fields. It
generalizes the dilations of flat spacetime.

Generalized Killing tensors of various types can also be generated
from Jacobi fields. In analogy to \eqref{AffineKT}, let
\begin{equation}
  \mathcal{K}_{ab} = \LieS g_{ab} .
  \label{KTApprox1}
\end{equation}
These objects exactly satisfy \eqref{KTDef} at $\gamma$, and
presumably approximate it near this point. It is straightforward to
write down other objects which also have this property. For example,
two (possibly identical) Killing-type Jacobi fields $\psi^a$ and
$\bar{\psi}^a$ can be used to define
\begin{equation}
  \mathcal{K}'_{ab} = \psi_{(a}
  \bar{\psi}_{b)} .
  \label{KTApprox2}
\end{equation}
This expression clearly generalizes to approximate Killing tensors
of any rank. Exact second-rank Killing tensors probably exist that
cannot be written in either of these forms, so it is unclear how
useful they are.

Very near $\gamma$, it is possible to approximate the Jacobi fields
explicitly. This will be especially useful in Sect.
\ref{Sect:ConsLaws} below, where a notion of gravitational current
is introduced with respect to a given vector field. Consider a
Taylor expansion of $\LieS g_{ab}$ in powers of $X^\sfa$. The first
two terms in this series are trivially obtained from
\eqref{JacobiAffine} and \eqref{InitialData}. Better approximations
involve third and higher derivatives of $\LieS g_{ab}$ in the limit
$x \rightarrow \gamma$. The lowest order interesting terms can be
found from \eqref{Del2Lieg} and \eqref{Del3Lieg}. Making use of
\eqref{Sig4Coinc}, the final results are that
\begin{equation}
  \fl \quad \LieS g_{ab} \simeq \sigma^{\sfa}{}_{a} \sigma^{\sfb}{}_{b} \big[ \LieS g_{\sfa \sfb} - \frac{1}{3} X^\sfc X^\sfd ( \LieS R_{\sfa \sfc \sfb \sfd}
  + \frac{1}{2} X^\sff \LieS \nabla_\sff R_{\sfa \sfc \sfb \sfd}) \big] + \Or(X^4) , \label{LieGExpandJacobi}
\end{equation}
and
\begin{eqnarray}
  \fl \quad \nabla_c \LieS g_{ab} \simeq - \frac{2}{3} \sigma^{\sfa}{}_{a} \sigma^{\sfb}{}_{b}
  \sigma^{\sfc}{}_{c} X^\sfd \big[ R_{\sfd \sfc
  (\sfa}{}^{\sff} \LieS g_{\sfb) \sff} + g_{\sff(\sfa} \LieS R_{\sfb) \sfd \sfc}{}^\sff  + \frac{3}{4}
  X^\sff  \nonumber
  \\
  \fl \qquad ~ \times \big( \frac{1}{3} g_{\mathsf{h} (\sfa} \LieS
  \nabla^{\mathsf{h}} R_{\sfb) \sfd \sff \sfc} - g_{\mathsf{h} \sfa } \LieS \nabla_\sfd R_{\sff (\sfb
  \sfc)}{}^{\mathsf{h}} - g_{\mathsf{h} \sfb} \LieS \nabla_\sfd
  R_{\sff (\sfa \sfc)}{}^{\mathsf{h}} \big) \big] + \Or(X^3).
  \label{LieGGradJacobi}
\end{eqnarray}
Lie derivatives here are evaluated at $\gamma$, so they only involve
$A_\sfa$, $B_{\sfa \sfb}$, $g_{\sfa \sfb}$, $R_{\sfa \sfb \sfc
\sfd}$, and its first two derivatives. The factors of
$\sigma^{\sfa}{}_{a}$ in front of these equations are used as a
convenient means for converting tensors at $\gamma$ into tensors at
$x$. It is perhaps more typical to use parallel propagators
$g^{\sfa}{}_{a}$ for this purpose \cite{PoissonRev}, although the
aforementioned simplicity of $\sigma^{\sfa}{}_{a}$ in normal
coordinates makes it an attractive alternative. There is very little
difference at low orders regardless. $\sigma^{\sfa}{}_{a}$ can be
freely interchanged with $-g^{\sfa}{}_{a}$ in
\eqref{LieGGradJacobi}. This is also possible in
\eqref{LieGExpandJacobi} when $B_{(\sfa \sfb)} = 0$.

Approximations like these are not useful over regions where the
curvature changes significantly, or on length scales approaching the
curvature radius. An alternative approach is to expand the various
bitensors built from $\sigma$ using its definition as an integral
along a geodesic. Simplifications can often be introduced by
ignoring all terms nonlinear in the Riemann tensor. A general method
for this type of weak-field procedure may be found in \cite{Synge,
deFelice}. Specific details involved with expanding the Jacobi
fields in this way will not be given here.

\section{Symmetries near a worldline}\label{Sect:GAC}

The Jacobi fields just discussed generalize the idea of a Killing
field or more general affine collineation in a normal neighborhood
of a given point. This is useful for some purposes, although it does
not have a very clear physical interpretation. The choice of origin
should presumably correspond to a preferred point, although there
are few of these that might arise in practice. It is often more
useful to base the idea of an approximate symmetry off of a given
timelike worldline rather than a single point. This could correspond
to the path of some observer. In some cases, the physical system
picks out preferred reference frames. A binary star system
experiencing no mass transfer can admit three center-of-mass frames
(rigorously defined in \cite{Dix70a, EhlRud,CM}), for example. Two
of these are associated with the individual stars, while the third
describes the system as a whole. There are also preferred observers
in most cosmological models. Expressing a system's dynamics in terms
of quantities associated with these frames has an obvious physical
interpretation. The distinction between approximate symmetries
defined with respect to a point versus a worldline is closely
analogous to the one between Riemann and Fermi normal coordinate
systems.

The concept of an observer here will be taken to mean a timelike
worldline $\Gamma$ together with a set of hypersurfaces $\Sigma(s)$
that foliate a surrounding worldtube $W$. It will be assumed that
each of these hypersurfaces is a normal neighborhood of the point
$\gamma(s)$ where it intersects the central worldline. Each of them
is therefore formed by a collection of radially-emanating geodesics
of (usually) finite length. The most typical examples would be the
past-directed null geodesics or the spacelike set orthogonal to
$\dot{\gamma}^\sfa = \rmd \gamma^\sfa/\rmd s$ at $\gamma(s)$. Other
choices are possible, however. Regardless, a worldline and foliation
together will be referred to as an observer's reference frame.

\subsection{A family of Jacobi fields}

Symmetries adapted to a particular frame can be constructed using a
one-parameter family of Jacobi fields $\psi^a(x,\gamma(s))$. Any
such family is fixed by specifying $A_\sfa(s)$ and $B_{\sfa\sfb}(s)$
as defined in \eqref{InitialData}. An optimal way of connecting
initial data between different points on $\Gamma$ therefore must be
found. Before considering this problem, it is first useful to
collapse the family of Jacobi fields into an ordinary vector field
$\xi^a(x)$. Let $\tau(x)$ be defined so as to identify which leaf of
the foliation includes an arbitrary point $x$ in the worldtube $W$.
More concisely, it always satisfies $x \in \Sigma(\tau(x))$. The
assumption that each hypersurface is a normal neighborhood of an
appropriate point on $\Gamma$ implies that $\tau$ is always
single-valued. Now set
\begin{equation}
  \xi^a (x) = \psi^a(x, \gamma(\tau(x))) .
  \label{XiDef}
\end{equation}
The generalized affine collineations (GACs) to be defined below will
be of this form for a particular class of families
$\psi^a(x,\gamma(s))$.

One potential application for a generalized symmetry constructed
using a particular frame is in the definition of quantities that
might be approximately conserved as one moves along $\Gamma$. As an
example, consider integrals of conserved stress-energy tensors
similar to (\ref{DixMomenta}). One might define the component of
momentum generated by a $\xi^a$ of the form \eqref{XiDef} to be
\begin{equation}
  \ItP_\xi(s) = \int_{\Sigma(s)} T^{a}{}_{b}
  \xi^b \rmd S_a . \label{PDef}
\end{equation}
The evolution of this quantity crucially depends on how the
parameters $A_\sfa(s)$ and $B_{\sfa \sfb}(s)$ in (\ref{InitialData})
are connected along $\Gamma$. It is well-known that $\ItP_\xi$ is
conserved if $\xi^a$ is Killing and no matter flows across the
boundary of the worldtube. In this case, initial data for the
one-parameter family of Jacobi fields must satisfy the Killing
transport (KT) equations on $\Gamma$:
\numparts
\begin{eqnarray}
    \mathrm{D}A_\sfa/\rmd s &= \dot{\gamma}^\sfb B_{\sfb \sfa}
    \label{KTA}
    \\
    \mathrm{D} B_{\sfa \sfb} / \rmd s &= - R_{\sfa \sfb
    \sfc}{}^{\mathsf{d}} \dot{\gamma}^\sfc A_{\mathsf{d}} .
    \label{KTB}
\end{eqnarray}
\endnumparts
If there exists an exact Killing vector $Y^a_{\mathrm{K}}$ such that
$A^\sfa =Y^\sfa_{\mathrm{K}}$ and $B^{\sfa \sfb} = \nabla^\sfa
Y^\sfb_{\mathrm{K}}$ at a given $s = s_0$, relations like these will
hold for all $s$. Furthermore, momenta $p^\sfa$ and $S^{\sfa \sfb}$
identified using (\ref{DixMomenta}) would satisfy
\begin{equation}
  0 = (\dot{p}_{\sfa} - \frac{1}{2} S^{\sfb \sfc} R_{\sfb \sfc \mathsf{d} \sfa}
  \dot{\gamma}^\sfc) Y^{\sfa}_{\mathrm{K}} + \frac{1}{2} (
  \dot{S}_{\sfa \sfb} - 2 p_{[\sfa} \dot{\gamma}_{\sfb]} )
  \nabla^{\sfa} Y^\sfb_{\mathrm{K}} . \label{Papapetrou}
\end{equation}
If there were a full complement of ten Killing vectors, all possible
versions of this expression would together be equivalent to the
Papapetrou equations. More generally, Papapetrou's result is only an
approximation. Any deviations can be understood using a general
10-parameter family of possibly approximate isometries. One might
expect these corrections to be minimized if $A_\sfa$ and $B_{\sfa
\sfb}$ always satisfy the KT equations even when no exact Killing
fields exist.
\begin{definition}\label{Def:GAC}
  Let a generalized affine collineation (GAC) $\xi^a(x)$ associated with a
  reference frame $\{\Gamma, \Sigma\}$ be derived from a family of
  Jacobi fields via \eqref{XiDef}. Individual elements of the family and their
  first derivatives satisfy the Killing transport equations \eqref{KTA} and \eqref{KTB} on $\Gamma$.
\end{definition}

Although this definition was motivated by the properties of
conserved momenta in very particular spacetimes, it also arises from
much more general (if less physical) arguments. Consider all
possible initial data for vectors built from Jacobi fields using
\eqref{XiDef}. It is reasonable to suppose that any GAC should be
exactly affine on $\Gamma$; i.e.
\begin{equation}
  \nabla_\sfa \LieX g_{\sfb \sfc} |_\Gamma = 0.
  \label{GACAffine}
\end{equation}
It can also be expected that $A_\sfa$ and $B_{\sfa \sfb}$ fix
$\xi^a$ and its first derivatives on $\Gamma$ just as they do for
$\psi^a$. Generalizing \eqref{InitialData}, let
\begin{equation}
  A^{\sfa}(s) = \xi^\sfa(\gamma(s)) , \qquad B^{\sfa \sfb}(s) =
  \nabla^{\sfa} \xi^\sfb(\gamma(s)) . \label{InitialDataXi}
\end{equation}

We start with the second of these constraints. Directly
differentiating (\ref{XiDef}) implies that
\begin{equation}
  \LieX g_{ab} = \LieS g_{ab} + 2 \dot{\psi}_{(a} \nabla_{b)} \tau ,
  \label{LieXivsPsi}
\end{equation}
where the Lie derivative with respect to $\psi^a(x, \tau)$ on the
right-hand side is understood (as usual) to involve only the first
argument of this vector field. It will be assumed that the foliation
is always sufficiently smooth that derivatives of $\tau$ remain
well-defined throughout $W$, and on $\Gamma$ in particular.
Evaluating \eqref{LieXivsPsi} on the central worldline requires
knowledge of $\dot{\psi}^a(\gamma,\gamma)$. Coincidence limits like
these are commonly denoted with brackets. For example,
\begin{equation}
  [\dot{\psi}^a](\gamma) = \lim_{x \rightarrow \gamma}
  \frac{ \partial}{\partial s} \dot{\psi}^a(x,\gamma(s)) .
  \label{CoincDef}
\end{equation}
The convention of using different fonts for indices referring to $x$
and $\gamma$ cannot be consistently applied in expressions like
this. No confusion should arise, however. Limits like
\eqref{CoincDef} are easily computed using Synge's rule \cite{Synge,
PoissonRev}. In this case,
\begin{eqnarray}
  [\dot{\psi}^a] &= \dot{\gamma}^\sfb [\nabla_\sfb \psi^a] \nonumber
  \\
   &= \dot{\gamma}^\sfb ( \nabla_\sfb [\psi^a] - \delta^b_\sfb [\nabla_b \psi^a]
  ) \nonumber
  \\
   &= \mathrm{D} A^a/\rmd s - \dot{\gamma}^\sfb B_{\sfb}{}^{a} .
  \label{DotPsi}
\end{eqnarray}
It follows that (\ref{InitialDataXi}) holds for all initial data iff
\eqref{KTA} is satisfied.

The other KT equation arises from enforcing \eqref{GACAffine}.
Noting (\ref{JacobiAffine}) and (\ref{KTA}), it must be true that
\begin{equation}
  [\nabla_a \dot{\psi}^b] = 0.
  \label{DelDotPsiReq}
\end{equation}
$[\ddot{\psi}^a]$ also has to vanish, although this term is equal to
$-\dot{\gamma}^b [ \nabla_b \dot{\psi}^a]$. Requiring
\eqref{DelDotPsiReq} is therefore sufficient. Using the same type of
procedure as in \eqref{DotPsi} shows that
\begin{equation}
  [\nabla_a \dot{\psi}_b] = \mathrm{D} B_{ab} / \rmd s+ R_{abc}{}^{d}
  \dot{\gamma}^c A_d . \label{DelDotPsi}
\end{equation}
Deriving this is straightforward other than noting that
\eqref{Del2Affine} -- although mentioned for exact affine
collineations -- also holds for any Jacobi field at its origin.
Regardless, the conclusion is that the second Killing transport
equation \eqref{KTB} ensures that $\nabla_\sfa \LieX g_{\sfb\sfc}$
vanishes everywhere on $\Gamma$.

Noting that the KT equations have the same significance for general
affine collineations as they do for ordinary Killing fields
\cite{Hall}, it easily follows that
\begin{theorem}\label{Thm:GACBasic}
  The class of all generalized affine collineations associated with
  a given reference frame forms a 20-dimensional group in four
  spacetime dimensions. Every GAC satisfies $\LieX \nabla = 0$ on
  $\Gamma$, and all exact affine collineations are members of this
  class.
\end{theorem}
This is closely related to theorem \ref{Thm:JacobiBasic}. It
strongly supports definition \ref{Def:GAC} and the intuitive
identification of GACs with approximate symmetries inside $W$.

At least in principle, finding GACs associated with a particular
reference frame is straightforward. Suppose that $A_\sfa(s_0)$ and
$B_{\sfa \sfb}(s_0)$ are given as initial data at some $\gamma_0 =
\gamma(s_0)$. The goal is then to determine the $\xi^a(x)$
satisfying \eqref{InitialDataXi} at the appropriate point. This is
done by first applying the KT equations to the given parameters
along $\Gamma$ from $\gamma_0$ to $\gamma(\tau(x))$. The geodesic
deviation equation (\ref{Jacobi}) is then integrated between this
latter point and $x$ using the initial conditions
(\ref{InitialData}). Both of these operations simply require finding
the solutions to well-behaved ordinary differential equations.
Alternatively, $\xi^a$ could also be obtained using the explicit
expression \eqref{JacobiFirstDer} together with the KT equations and
\eqref{XiDef}.

Our prescription for generalizing arbitrary affine collineations may
appear somewhat awkward. Killing transport equations are being
applied along $\Gamma$, while the Jacobi equation is used on
geodesics intersecting that worldline. These two procedures are not
as different as they might appear. Trying to use Killing transport
everywhere would generically lead to inconsistencies. Derivatives of
the field expected from the KT equations would not usually match the
derivatives computed from $\xi^a$ itself. Only the tangential
components of these derivatives can be consistently fixed by
integrating ordinary differential equations along a collection of
radial geodesics. Weakening the Killing transport equations to take
this into account exactly reproduces the geodesic deviation
equation. This may be seen by rewriting \eqref{Jacobi} as a pair of
first order differential equations on geodesics connecting $x$ to
$\gamma(\tau(x))$. Denote the unit tangent vector to one such
geodesic by $u^a(l)$. Also set $\hat{A}^a = \psi^a$ and
$\hat{B}_{ab} = \nabla_a \psi_b$ everywhere. It is then
straightforward to show that \numparts
\begin{eqnarray}
  \mathrm{D} \hat{A}^a/\rmd l &= u^b \hat{B}_{ba}
  \\
  u^a \mathrm{D} \hat{B}_{ab}/\rmd l &= - R_{abc}{}^{d} u^a u^d
  \hat{A}_d.
\end{eqnarray}
\endnumparts
The first of these equations has exactly the same form as
\eqref{KTA}, while the second is essentially \eqref{KTB} contracted
with $u^a$. Killing and Jacobi transport are therefore very closely
related operations. The latter does not uniquely propagate
$\hat{B}_{ab}$ from given initial data, so it is weaker. These
remarks also clarify in what sense Jacobi fields or GACs approximate
\eqref{AffineDef} or \eqref{Del2Affine}.

\subsection{Special cases and properties of GACs}

From a physical perspective, momenta like \eqref{PDef} should be
definable even in the absence of any exact isometries. This is most
conveniently done with a particular class of GACs that generalize
only the Killing fields. In analogy to the Killing-type Jacobi
fields discussed in Sect. \ref{Sect:SpecialJacobi}, suppose that
$B_{(\sfa \sfb)}$ vanishes on at least one point of $\Gamma$. It
immediately follows from \eqref{KTB} that it must actually vanish
everywhere. The Killing-type GACs therefore form a 10-dimensional
group of vector fields satisfying
\begin{equation}
  \Lie_{\xi_\mathrm{K}} g_{\sfa \sfb} |_\Gamma = 0,
  \label{GACKilling}
\end{equation}
as well as \eqref{GACAffine}. They may be thought of as generalizing
the Poincar\'{e} symmetries of flat spacetime near a given observer.

It also possible to single out GACs that are approximately
homothetic in the sense that
\begin{equation}
  \Lie_{\xi_{\mathrm{H}}} g_{\sfa \sfb} |_\Gamma = 2 c g_{\sfa
  \sfb}. \label{GACHomothetic}
\end{equation}
This requires setting $B_{(\sfa \sfb)} = c g_{\sfa \sfb}$. While
always possible, the remaining components of the initial data cannot
be explicitly solved for except in the case when $\Gamma$ is a
geodesic. It then self-consistent to choose $B_{[\sfa \sfb]}=0$. The
obvious way of doing this is to normalize $\dot{\gamma}^\sfa$ to
unity and set
\begin{equation}
  A_\sfa = c (s - \bar{s} ) \dot{\gamma}_\sfa
\end{equation}
for some constant $\bar{s}$. It is easily verified that the given
parameters satisfy the KT equations. One homothetic-type GAC
associated with an affinely parameterized geodesic therefore has the
form
\begin{equation}
  \xi^a_{\mathrm{H}} = c \big[ \sigma^a + (\tau - \bar{s}) H^{a}{}_{\sfa}
  \sigma^{\sfa}{}_{\sfb} \dot{\gamma}^\sfb \big] .
\end{equation}
As usual, Killing-type Jacobi fields may be added to this without
spoiling \eqref{GACHomothetic}. It should also be emphasized that
homothetic-type GACs are not restricted to geodesic frames. This is
just the case where closed-form solutions of the KT equations can be
obtained by inspection.

Many of the properties derived for Jacobi fields in Sect.
\ref{Sect:Symmetries} and the appendix can be carried over at least
partially for the GACs. For example,
\begin{equation}
  \LieX \sigma^\sfa = \LieS \sigma^\sfa  = 0  \label{GACLie1}
\end{equation}
if the arguments are of the form $(x, \gamma(\tau(x)))$ and $\xi^a$
and $\psi^a$ are related via \eqref{XiDef}. This may be interpreted
as stating that spatial Fermi coordinates are preserved under flows
generated by $\xi^a$. Since the hypersurfaces $\Sigma(s)$ can be
described as a set of geodesics intersecting $\gamma(\tau(x))$, it
will always be true that $\sigma^a \nabla_a \tau =0$. This means
that
\begin{theorem}\label{Thm:GACExactSyms}
  Given a general GAC $\xi^a$, $\LieX \sigma^\sfa = \LieX \sigma^a
  =0$ when the arguments of these equations are as in
  \eqref{GACLie1}. Killing-type GACs $\xi^a_{\mathrm{K}}$ also satisfy $\Lie_{\xi_{\mathrm{K}}} \sigma = \Lie_{\xi_{\mathrm{K}}} \sigma_\sfa =
  0$ with the same restriction.
\end{theorem}

The identity \eqref{LieSig6} serves to constrain Lie derivatives of
the metric with respect to Jacobi fields. A direct analog of this
equation for an arbitrary GAC would involve an additional term.
Despite this, contracting the result with $\sigma^b$ leads to the
simple conclusion
\begin{equation}
  \sigma^a \sigma^b \LieX g_{ab} = 2 \sigma^\sfa \sigma^\sfb B_{(\sfa
  \sfb)} .
\end{equation}
``Purely radial'' components of $\LieX g_{ab}$ therefore vanish for
Killing-type GACs.

Many other results can be carried over in similar ways. One that is
of particular interest is the behavior of $\LieX g_{ab}$ or
$\nabla_a \LieX g_{bc}$ near $\Gamma$. As with the Jacobi fields, it
is possible to see how close a GAC comes to being affine as its
reference worldline is approached. Analogs of
\eqref{LieGExpandJacobi} and \eqref{LieGGradJacobi} may be obtained
using expansions like \eqref{LieXivsPsi} and the identity
\eqref{Del2General}. Simplifying terms with the Killing transport
equations, the lowest order correction to \eqref{LieGExpandJacobi}
is
\begin{equation}
  \Lie_{(\xi-\psi)} g_{ab} \simeq \frac{2}{3} \sigma^{\sfa}{}_{a}
  \sigma^{\sfb}{}_{b} \nabla_{(\sfa} \tau g_{\sfb) \sff}
  \dot{\gamma}^{\mathsf{h}}
  X^\sfc X^\sfd \LieX R_{\mathsf{h} \sfc \sfd}{}^{\sff} + \Or(X^3).
  \label{LieGExpandGAC}
\end{equation}
Similarly, the first interesting change to \eqref{LieGGradJacobi}
has the form
\begin{eqnarray}
  \fl \qquad \nabla_a \Lie_{(\xi-\psi)} g_{bc} \simeq - \frac{4}{3} \sigma^{\sfa}{}_a
  \sigma^{\sfb}{}_b \sigma^{\sfc}{}_c X^\sfd \dot{\gamma}^\sff
  \big[ g_{\mathsf{h} (\sfb} \nabla_{\sfc)} \tau (\delta^{\mathsf{l}}_\sfa - \dot{\gamma}^{\mathsf{l}} \nabla_\sfa \tau) \LieX R_{\sff (\mathsf{l} \sfd)}{}^{\mathsf{h}} \nonumber
  \\
  \qquad ~  + \frac{1}{2} \nabla_\sfa \tau (R_{\sfd \sff
  (\sfb}{}^{\mathsf{h}} \LieX g_{\sfc) \mathsf{h}} - g_{\mathsf{h} (\sfb} \LieX R_{\sfc) \sff \sfd}{}^{\mathsf{h}} ) \big] + O(X^2).
  \label{LieGGradGAC}
\end{eqnarray}
As before, the magnitudes of these terms depend on how close $\xi^a$
is to being a symmetry of the Riemann tensor on the observer's
worldline.

\section{Mechanics and conservation laws}
\label{Sect:ConsLaws}

It has already been remarked that one of the main applications of
exact symmetries in physics is to the formulation of conservation
laws. These take several forms. Perhaps the most basic are those
associated with a spacetime's geodesics. It is sometimes possible
for such curves to be at least partially parameterized by a number
of constants associated with geometric symmetries. More
interestingly, conservation laws can also be associated with
extended matter distributions. Assuming only that stress-energy
tensors satisfy
\begin{equation}
  \nabla_a T^{ab} = 0 \label{StressCons}
\end{equation}
tends to lead to the definition of slowly-varying parameters like
those discussed in connection with \eqref{DixMomenta} and
\eqref{PDef}. The situation becomes much more interesting in full
general relativity. Einstein's equation implies stress-energy
conservation, although it also connects symmetries of the geometry
to those of the matter distribution (and vice versa). This allows
the introduction of exact conservation laws in arbitrary spacetimes.

\subsection{Geodesics}

It is well-known that any Killing fields that may exist provide
first integrals of the geodesic equation. These can be used both to
derive and parameterize the geodesics of a given spacetime. While
less commonly discussed, similar quantities can also be associated
with other kinds of symmetries \cite{GeodesicConsts,KatzinGeo}.
Unlike in the Killing vector case, the presence of more general
collineations sometimes implies the existence of interesting
conserved quantities that are not linear in the geodesic's
four-velocity. The curve's affine parameter can also appear
explicitly. As a direct calculation will easily verify, two
constants associated with an exact affine collineation
$Y_{\mathrm{A}}^a$ are \numparts\label{GeoConsts}
\begin{eqnarray}
    C_1 &= \dot{y}^a \dot{y}^b \Lie_{Y_{\mathrm{A}}} g_{ab}
    \label{GeoConst1}
    \\
    C_2 & = \dot{y}_a Y^a_{\mathrm{A}} - \frac{1}{2} l C_1 .
    \label{GeoConst2}
\end{eqnarray}
\endnumparts
These quantities remain fixed along any affinely-parameterized
geodesic $y(l)$. The first becomes degenerate if $Y^a_{\mathrm{A}}$
is Killing. $C_2$ then reduces to the standard conserved quantity
associated with a Killing field. Other constants can sometimes be
written down by combining $Y^a_{\mathrm{A}}$ with an exact Killing
tensor \cite{GeodesicConsts, GeoExamples}. Such constructions will
not be discussed here.

Consider instead expressions like those just given with
$Y_{\mathrm{A}}^a(x)$ replaced by some Jacobi field
$\psi^a(x,\gamma)$. We then have
\begin{eqnarray}
  \dot{C}_1 = \frac{\rmd C_1}{\rmd l} = -\frac{2}{l} \frac{\rmd C_2}{ \rmd l} = \dot{y}^a \dot{y}^b \dot{y}^c \nabla_{(a} \LieS
  g_{bc)} .
\end{eqnarray}
The tangent vectors are proportional to $\sigma^a(y,\gamma)$ for the
special case of geodesics passing through $\gamma$. It then follows
from \eqref{LieSig9} that both $C_1$ and $C_2$ remain conserved
along all such trajectories. Each Jacobi field generates exact
geodesic constants in this way. In terms of the initial data
$A_\sfa$ and $B_{\sfa \sfb}$, these have the values
\begin{eqnarray}
  C_1  &= 2 \dot{y}^\sfa \dot{y}^\sfb B_{(\sfa \sfb)}, \qquad
  C_2 &= \dot{y}^\sfa A_\sfa \label{GeoConstInit}
\end{eqnarray}
when the parameter $l$ is chosen to vanish at $\gamma$. It is clear
from this that $B_{[\sfa \sfb]}$ is irrelevant. Multiple Jacobi
fields may therefore generate the same constants on a particular
curve.

Exact affine collineations generalize these results by also applying
to non-radial geodesics. Expansions like \eqref{LieGGradJacobi} can
be used to derive how close general Jacobi fields come to this
ideal. To lowest nonvanishing order,
\begin{equation}
  \dot{C}_1 \simeq - \frac{4}{3} (\dot{y}^a \sigma^{\sfa}{}_{a}) (\dot{y}^b \sigma^{\sfb}{}_{b}) (\dot{y}^c
   \sigma^{\sfc}{}_{c}) X^\sfd R_{\sfd \sfb
  \sfc}{}^{\sff} \LieS g_{\sfa \sff} + \Or(X^2) . \label{Cdot}
\end{equation}
This term vanishes if $B_{(\sfa \sfb)}=0$, so the Killing-type
Jacobi fields usually provide more accurate ``conservation laws''
for arbitrary geodesics near $\gamma$. In these cases,
$\dot{C}_1(l)$ scales like $(X/\mathcal{R})^2/\mathcal{R}$, where
$\mathcal{R}$ is a curvature radius. This is a worst-case estimate.
$C_1$ and $C_2$ will probably vary much more slowly if there is a
physical sense in which the system is approximately symmetric. Rates
of change for $C_1$ or $C_2$ can also be constrained using
identities like \eqref{LieSig7}. This effectively restricts how much
these parameters can vary as a geodesic moves away from $\gamma$.
More of their changes tend to occur as $y(l)$ moves across rather
than with the radial geodesics.

Parameters like $C_1$ and $C_2$ can also be defined with respect to
a GAC $\xi^a$. These should remain approximately conserved for
geodesics near an observer's worldline rather than geodesics near a
point. They are exact constants for curves passing through $\Gamma$
along the reference foliation. This can be seen by using
\eqref{LieXivsPsi} to show that
\begin{equation}
  \dot{y}^a \dot{y}^b \dot{y}^c \nabla_a \LieX g_{bc} =   \dot{y}^a \dot{y}^b \dot{y}^c \nabla_{a} \LieS g_{bc}
\end{equation}
when $\dot{y}^a \nabla_a \tau =0$. Values of $C_1$ and $C_2$ here
are the same as in \eqref{GeoConstInit} if the quantities in that
equation are evaluated at the point where $y$ intersects $\Gamma$.
This might be useful in a coordinate system constructed from some
collection of GACs. Alternatively, it can be viewed as a
generalization of standard results near a given observer.

Further methods of parameterizing geodesics may be found by
generalizing the constants associated with higher-rank Killing
tensors. Given any exact Killing tensor $K_{a_1 \cdots a_n} =
K_{(a_1 \cdots a_n)}$, the scalar
\begin{equation}
  C_K = K_{a_1 \cdots a_n} \dot{y}^{a_1} \cdots \dot{y}^{a_n}
  \label{CK}
\end{equation}
is conserved along all geodesics $y(l)$. An analog of this quantity
for the approximate Killing tensor \eqref{KTApprox1} is exactly
$C_1$ defined above. Something more interesting can be generated by
substituting \eqref{KTApprox2} into \eqref{CK}. In the second-rank
case, one may define
\begin{equation}
  C_K = (\dot{y}_a \psi^a_{\mathrm{K}}) (\dot{y}_b \bar{\psi}^b_{\mathrm{K}}) \label{CKEx}
\end{equation}
for some Killing-type Jacobi fields $\psi^a_{\mathrm{K}}$ and
$\bar{\psi}^a_{\mathrm{K}}$. This is easily generalized to involve
an arbitrary number of products, although it is only interesting to
consider linearly independent collections of Jacobi fields. The
maximum number of useful products is therefore ten. All of these can
be generated just from individual terms of the form $\dot{y}_a
\psi^a_{\mathrm{K}}$. These are interpreted as approximate constants
associated with objects that are nearly first-rank Killing tensors
(i.e. Killing vectors). Everything of interest here can therefore be
derived from the behavior of
\begin{equation}
  C_3 = \dot{y}_a \psi^a_{\mathrm{K}} = C_2 + \frac{1}{2} l C_1.
  \label{C3}
\end{equation}
Although this depends only on the two approximate constants defined
before, it may be interpreted as an additional useful parameter. It
has the interesting property that
\begin{equation}
  \dot{C}_3 = \frac{1}{2} C_1 . \label{C3Dot}
\end{equation}
Time derivatives do not appear on the right hand side. Consider the
special case of a geodesic that passes through $\gamma$. Since the
Jacobi field was assumed to be Killing at its origin, $C_1 = 0$.
This is true everywhere, so $C_3$ also remains fixed along the
entire geodesic. It actually coincides with $C_2$ in this case.

Differences arise when considering non-radial geodesics. It was
remarked above that there was a sense in which $C_1$ and $C_2$ only
varied due to non-radial components of $\dot{y}^a$. This type of
statement can be made much more precise for $C_3$. Let
\begin{equation}
  \dot{y}^a(l) = u_{||} (l) \sigma^a(y(l),\gamma) + u^a_\bot(l) ,
\end{equation}
where $u^{[a}_\bot \sigma^{b]} =0$. Now \eqref{LieSig6},
\eqref{GeoConst1}, and \eqref{C3Dot} show that
\begin{equation}
  \dot{C}_3 = \frac{1}{2} u_\bot^a u_\bot^b \LieS g_{ab} .
  \label{C3DotBeta}
\end{equation}
This result is exact for all geodesics $y(l)$. There are many cases
where $u_\bot^a$ becomes vanishingly small as $l \rightarrow \pm
\infty$ (when the geodesic exists for these parameter values), so
\eqref{C3DotBeta} provides a strong restriction on how much $C_3$
can vary in any given situation. Very near $\gamma$,
\eqref{LieGExpandJacobi} can be used to show that
\begin{equation}
  \dot{C}_3 \simeq - \frac{1}{6} (u_\bot^a \sigma^{\sfa}{}_{a}) (u_\bot^b \sigma^{\sfb}{}_{b})
  X^\sfc X^\sfd \LieS R_{\sfa \sfc \sfb \sfd} + \Or(X^3) .
\end{equation}
The lowest order contributions here scale like
$(X/\mathcal{R})^2/\mathcal{R}$. This is similar to the result
expected for $\dot{C}_1$ when computed using a Killing-type Jacobi
field.

\subsection{Extended matter
distributions}\label{Sect:ExtendedMatter}

From a physical perspective, it is often more interesting to
consider possibly approximate integrals of the equations of motion
describing an extended matter distribution rather than a pointlike
test particle. Suppose that this matter is modeled by a conserved
stress-energy tensor $T^{ab}$. Contracting it with any exact Killing
vectors that may exist yields conserved currents. These are
equivalent to some subset of the typical laws of linear and angular
momentum conservation known in flat spacetime. More generally,
\eqref{StressCons} shows that
\begin{equation}
  \nabla_a ( T^{a}{}_{b} Y^b ) = \frac{1}{2} T^{ab} \Lie_Y g_{ab} .
\end{equation}
This holds for any vector field $Y^a$, although it is convenient to
assume that it is a Killing-type GAC. The source term on the
right-hand side may then be considered small near $\Gamma$. This
therefore serves as an approximate conservation law. As long as
there is no matter flow through $\partial \Sigma$, quantities like
\eqref{PDef} might be expected to vary slowly in time.


Stress-energy conservation can be seen as a consequence of the
diffeomorphism invariance of a system's underlying action.
Constructions that are based on it are therefore useful in many
theories of gravity besides general relativity. They can also hold
for test bodies in fixed background geometries. This generality is
quite restrictive. Much more can be said if the full Einstein
equation is assumed to hold. Symmetries in the geometry are then
related to symmetries in the matter fields. The presence of an exact
Killing field $Y^a_{\mathrm{K}}$ that also satisfies
$\Lie_{Y_\mathrm{K}} T_{ab} = 0$ allows many interesting results to
be proven regarding the momenta $p_\sfa$ and $S_{\sfa \sfb}$ defined
in \eqref{DixMomenta}. For example, the net force and torque on a
body can be written explicitly in terms of the Killing field and its
first derivative at a point. If $Y^a_{\mathrm{K}}$ is timelike, a
body's center-of-mass can be shown to follow one of its orbits.
Furthermore, mass centers always lie on the central axis of
cylindrically symmetric spacetimes \cite{SchattStreub1,
SchattStreub2}.

Momenta defined in terms of $T^{a}{}_{b} \xi^b$ are useful for many
purposes, although they are not conserved in the absence of exact
Killing fields. Determining how they vary over time requires
detailed knowledge of a body's internal structure. Alternative
definitions for the linear and angular momenta of an extended body
arise when using the full Einstein equation rather than just
\eqref{StressCons}. Taking the trace of Ricci's identity and
rearranging terms shows that
\begin{equation}
  R^{a}{}_{b} Y^b = \frac{1}{2} ( g^{ac} g^{bd} - g^{ab} g^{cd} )
  \nabla_b \Lie_Y g_{cd} + \nabla_b \nabla^{[a} Y^{b]} .
  \label{RicciId}
\end{equation}
This holds for any vector field $Y^a$. Note that the second term on
the right-hand side is always conserved. It follows that
\begin{eqnarray}
  \nabla_a [ 2 (T^{a}{}_{b} Y^b -\frac{1}{2} Y^a T^{b}{}_{b}) + j^a_Y] = 0, \label{AffineConsLaw}
\end{eqnarray}
where the ``gravitational current'' $j^a_Y$ associated with $Y^a$
has been defined by
\begin{equation}
  j^a_Y = \frac{1}{8 \pi} ( g^{ab} g^{cd} - g^{ac} g^{bd} )
  \nabla_b \Lie_Y g_{cd} .
  \label{GravCurrent}
\end{equation}
It clearly vanishes if $Y^a$ is an exact affine collineation. This
is not the only case where the current's contribution to
\eqref{AffineConsLaw} disappears. Using the Bianchi identity,
\begin{equation}
  \nabla_a j^a_Y = - \frac{1}{8\pi} g^{ab} \Lie_Y R_{ab}.
  \label{divJ}
\end{equation}
Any vector field satisfying $g^{ab} \Lie_Y R_{ab} =0$ will therefore
generate conserved matter currents involving only $T^{a}{}_{b} Y^b -
Y^a T^{b}{}_b/2$. That such ``contracted Ricci collineations''
generate conservation laws for matter distributions has been noted
before in \cite{KatzinLevine, Collinson}.

The viewpoint here will be to apply \eqref{RicciId} and
\eqref{AffineConsLaw} with $Y^a$ replaced by some approximate
symmetry.
\begin{definition}
    Fix some family of Jacobi fields $\psi^a(x,\gamma(s))$ that generates a GAC
    via \eqref{XiDef}. Define the generalized Komar momentum
    $\ItP^*_\psi$ associated with these fields by
    \begin{equation}
        \ItP^*_\psi(s) = \frac{1}{8\pi} \oint_{\partial \Sigma(s)} \nabla^{[a}
        \psi^{b]} \rmd S_{ab} . \label{PStarDef}
    \end{equation}
\end{definition}
As the name suggests, this has the same form and interpretation as a
Komar integral. It is convenient to assume that $s$ is a fixed
parameter for the purpose of evaluating the derivative in
\eqref{PStarDef}. Directly using a GAC in place of $\psi^a$ would
add a dependence on the reference frame. Note that no such
distinctions had to be made for the $\ItP_\xi$ defined in
\eqref{PDef}. Applying Stokes' theorem together with \eqref{RicciId}
and \eqref{GravCurrent} shows that
\begin{equation}
  \ItP^*_\psi =  \int_\Sigma \big[ 2 (T^{a}{}_{b} \psi^b - \frac{1}{2}
  \psi^a T^{b}{}_{b} ) + j^a_\psi \big] \rmd S_a . \label{PStar2}
\end{equation}
There is a well-defined sense in which changes in this quantity are
determined by a combination of ``gravitational wave'' and matter
fluxes across the boundary $\partial \Sigma$. In this
interpretation, the amount of $\ItP^*_\psi$ carried away from a
system via gravitational waves vanishes if the GAC associated with
$\psi^a$ is an affine collineation.

The 20-parameter family of scalars $\ItP^*_\psi$ is intended to
define the linear and angular momenta of an extended body. This is
at least the interpretation for the 10-parameter subset satisfying
$B_{(\sfa \sfb)}=0$. As in \eqref{DixMomenta}, it is possible to
write these momenta in the more conventional form of tensor fields
on $\Gamma$. Let
\begin{equation}
  \ItP^*_\psi = p_\sfa^* A^\sfa + \frac{1}{2} S^*_{\sfa \sfb}
  B^{\sfa \sfb}\label{KomarMoment}
\end{equation}
for all $A_\sfa$ and $B_{\sfa \sfb}$. As written, the generalized
angular momentum $S_{\sfa\sfb}^*$ needn't have any particular index
symmetries. Non-Killing Jacobi fields generate the symmetric
components of this tensor, although such generality isn't necessary.
Varying among all combinations of initial data completely recovers
$p^*_\sfa$ and $S^*_{\sfa \sfb}$. Direct expressions can also be
obtained with the use of \eqref{JacobiFirstDer}. Continuing to work
with the less explicit form \eqref{KomarMoment}, rates of change of
the tensor momenta may easily be extracted from $\dot{\ItP}^*_\psi$.
Using the KT equations \eqref{KTA} and \eqref{KTB},
\begin{equation}
  \dot{\ItP}^*_\psi = ( \dot{p}^*_\sfa - \frac{1}{2} S^*_{\sfb \sfc}
  R^{\sfb \sfc}{}_{\sfd \sfa} \dot{\gamma}^\sfd ) A^\sfa +
  \frac{1}{2} ( \dot{S}^*_{\sfa \sfb} + 2 \dot{\gamma}_\sfa p^*_\sfb) B^{\sfa \sfb}.
\end{equation}
This is closely analogous to \eqref{Papapetrou}. The left-hand side
is parameterized entirely by $A_\sfa$ and $B_{\sfa \sfb}$, so
varying these quantities determines all of the corrections to the
Papapetrou equations.

Two definitions have now been suggested for the momenta of an
extended body. The first -- summarized by \eqref{DixMomenta} and
\eqref{PDef} -- is closely related to the one given by Dixon
\cite{Dix70a, Dix74, Dix79}. It is well-adapted to the construction
of multipole moments for $T^{ab}$ that intrinsically take into
account stress-energy conservation. Mass centers defined from these
momenta are known to have most of the properties one might expect
\cite{SchattStreub1,CM}. The boundary of the worldtube $W$ isn't
important as long as it lies outside of the matter distribution
under discussion. There is no vacuum momentum under this definition.
Unfortunately, there does not appear to be any exact analog of
Gauss' law either. The momenta of a matter distribution (and changes
to it) must be computed by integrating over 3-volumes. The
generalized Komar integrals defined by \eqref{PStarDef} have
complementary characteristics. Their main advantage is in having a
direct interpretation analogous to Gauss' law. Changes in the
component of an isolated body's momentum generated by $\psi^a$ only
depends on the gravitational flux $j^a_\psi$ passing through the
surface $\partial W$. The mass and angular momenta expected from
this definition also agree with commonly-accepted notions at least
in appropriately symmetric spacetimes. It is potentially problematic
that $\ItP^*_\psi$ includes what is effectively a vacuum energy.
These scalars usually depend on the spatial extent of $W$ even when
its boundary lies far outside of any matter distribution. This can
make it difficult to neatly separate the properties of disjoint
matter distributions, although similar situations are found even in
ordinary electromagnetism. It might be conceptually simpler to
extend $\partial W$ to infinity, although it is unlikely that all of
the bitensors used here would remain well-defined. There might also
be convergence problems. Related concepts presented in
\cite{EnergyConference} could be more useful for defining momenta
over very large regions.

Some insight into the behavior of the $\ItP^*_\psi$ defined here can
be gained by computing it for very small spheres. To be specific,
let $C(r,s)$ be the closed 2-surface on $\Sigma(s)$ satisfying
$X^\sfa X_\sfa = r^2$ for some $r>0$. This is effectively a sphere
of proper radius $r$ centered at the point $\gamma(s)$. It follows
from \eqref{GACAffine} and \eqref{GravCurrent} that $j^a_\psi$ is
negligible for small radii in the presence of matter. The momentum
inside $C$ is approximately
\begin{equation}
  P^*_\xi \simeq \frac{8\pi}{3} r^3 (T^{\sfa}{}_{\sfb} A^\sfb -
  \frac{1}{2} A^{\sfa} T^{\sfb}{}_{\sfb}) \nabla_\sfa \tau + \Or(r^4) .
\end{equation}
It is perhaps more interesting to consider regions that are locally
devoid of matter. These can be understood from the behavior of the
gravitational current. Its approximate behavior near $\Gamma$ is
easily calculated from \eqref{LieGGradJacobi} and
\eqref{GravCurrent}. To lowest nontrivial order,
\begin{eqnarray}
  \fl \qquad \qquad j^a_\psi \simeq - \frac{1}{8 \pi} \sigma^{a}{}_{\sfa} X^\sfb \big[
  \LieS R^{\sfa}{}_{\sfb} + \frac{1}{3} \big( g^{\sfa \sfc} R^{\sfd}{}_{\sfb} + 2 R_{\sfb}{}^{\sfc \sfa \sfd} \big) \LieS
  g_{\sfc \sfd} \big] + \Or(X^2). \label{GravCurrentEst}
\end{eqnarray}
This vanishes in vacuum for Killing-type Jacobi fields. A little
more calculation finds the same conclusion at order $X^2$ as well.
This contrasts sharply with other quasilocal notions of vacuum
momentum in general relativity. An extensive review of these
concepts may be found in \cite{EnergyRev}. As remarked there, the
energy contained in small spheres has been calculated using several
different definitions. The generic result is that it is proportional
to the Bel-Robinson tensor, and scales like $r^5$. If this were true
for the definition suggested here, terms quadratic in the curvature
would appear at order $X^2$ in the current. These are not found.
Vacuum momenta should really only be associated with Killing-type
symmetries, so the relevant currents defined here decrease at least
as fast as $r^6$ as $r \rightarrow 0$. Other definitions in the
literature find more energy in very small spheres. It is not clear
how to interpret this, although it might have interesting
consequences for the use of near zones and related concepts
connected to the mechanics of compact bodies.


\section{An example: gravitational plane waves}\label{Sect:Example}

The discussion so far has mainly focused on the behavior of
generalized symmetries near the point or worldline used to construct
them. With the exception of general identities like \eqref{LieSig6},
it has not been clear what happens to these vector fields far away
from their origins. It is therefore useful to consider an example.
Given \eqref{JacobiFirstDer}, the Jacobi fields can all be
calculated simply by differentiating the world function. This makes
it convenient to consider spacetimes where $\sigma$ is known
exactly. Essentially the only examples of this type are
\textit{pp}-waves or assorted cosmological models (see
\cite{WorldFunction} and references cited therein).

In the interest of understanding the generalized Komar momenta
\eqref{PStarDef} in a vacuum spacetime, only \textit{pp}-waves will
be considered here. Coordinates may be introduced such that the
metric satisfies
\begin{equation}
  \rmd s^2 = - 2 \rmd u \rmd v + a(u) \rmd x^2 + b(u) \rmd y^2 .
  \label{ppGeneral}
\end{equation}
It can then be shown that one-half of the geodesic distance between
points with coordinates $(u,v,x,y)$ and $(\mathsf{u}, \mathsf{v},
\mathsf{x}, \mathsf{y})$ is given by \cite{Gunther,Friedlander}
\begin{equation}
  \sigma = \frac{1}{2} [ \alpha (u,\mathsf{u}) (x-\mathsf{x})^2 + \beta (u,\mathsf{u}) (y-\mathsf{y})^2 ] -
  (u-\mathsf{u}) (v-\mathsf{v}), \label{SigmaExact}
\end{equation}
where
\begin{equation}
  \alpha(u,\mathsf{u}) = \frac{u-\mathsf{u}}{\int_{\mathsf{u}}^{u} a^{-1}(w) \rmd w} ; \quad \beta(u,\mathsf{u}) = \frac{u-\mathsf{u}}{\int_{\mathsf{u}}^{u} b^{-1}(w) \rmd
  w}.
\end{equation}
It is clear by inspection that $\partial/\partial x$,
$\partial/\partial y$, and $\partial/\partial v$ are exact Killing
vectors. They are not the only ones. All nontrivial spacetimes in
this class admit between five and seven linearly-independent Killing
fields.

Most \textit{pp}-waves are effectively ``null dust'' solutions of
Einstein's equation, although there are vacuum examples as well. One
of these is given by
\begin{equation}
  a(u) = \cos^2 (\lambda u) ; \quad b(u) = \cosh^2 (\lambda u)
 , \label{GravWave}
\end{equation}
where we assume that $|\lambda u|< \pi/2$ in order to avoid the two
coordinate singularities. This represents a simple plane-fronted
gravitational wave with amplitude $\lambda$. The only non-vanishing
components of the curvature are
\begin{equation}
  C_{uxu}{}^{x} = C_{uyu}{}^{y} = \lambda^2.
\end{equation}
It is trivial to modify this spacetime to be flat for (say) $u<0$
\cite{PlaneWaves}, although impulsive waves of this type will not be
discussed here.

Given \eqref{SigmaExact}, it is straightforward to explicitly
compute the Jacobi propagators $H^{a}{}_{\sfb}
\sigma^{\sfb}{}_{\sfa}$ and $H^{a}{}_{\sfa} \sigma_\sfb$ using
\eqref{HDef}. This can be done for any choice of $a(u)$ and $b(u)$,
although we will specialize to the case defined by \eqref{GravWave}.
The results are not particularly enlightening to write down in
detail, although they have some interesting consequences. First, all
Jacobi fields are found to be exactly Killing if their first
derivatives vanish at the base point (denoted as usual with
sans-serif font). The Jacobi fields $H^{a}{}_{[\mathsf{x}}
\sigma_{\mathsf{v}]}$ and $H^{a}{}_{[\mathsf{y}}
\sigma_{\mathsf{v}]}$ are also Killing. This identifies six
independent Killing fields. It also implies that the linear
gravitational momentum $p^*_\sfa $ defined in \eqref{KomarMoment}
must vanish. This can be taken to imply (unsurprisingly) that the
gravitational wave has zero rest mass: $|p^*|^2=0$.

There remain four non-Killing Jacobi fields with skew-symmetric
$B_{\sfa \sfb}$. The one associated with spatial rotations in the
$x-y$ plane is relatively simple to write down when $\mathsf{u}=0$:
\begin{eqnarray}
  \fl \qquad \psi_{\mathsf{xy}}^a = 2 H^{a}{}_{\mathsf{[x}} \sigma_{\mathsf{y}]} =
  -(y-\mathsf{y})
  \left( \frac{ \tan(\lambda u) }{ \tanh(\lambda u) } \right)
  \frac{\partial}{\partial x} + (x - \mathsf{x}) \left( \frac{ \tanh(\lambda u) }{
  \tan(\lambda u) } \right) \frac{\partial}{\partial y}
  \nonumber
  \\
  \qquad \qquad ~  + \lambda ( x -\mathsf{x})
  (y - \mathsf{y}) \left( \frac{ \tanh(\lambda u) - \tan (\lambda u) }{ \tan(\lambda u) \tanh(\lambda u) } \right) \frac{\partial}{\partial
  v} .
\end{eqnarray}
This clearly reduces to its expected form when $u \rightarrow
\mathsf{u}$. The degree to which it succeeds in being a genuine
Killing field may be estimated by noting that
\begin{equation}
    \fl \qquad |\mathcal{L}_{\psi_{\mathsf{xy}}} g_{ab}|^2 = \left( \frac{\cos(2 \lambda u) + \cosh(2 \lambda u) -2 }{\sqrt{2} \sin(\lambda u) \sinh(\lambda
    u)} \right)^2 \simeq \frac{8}{9} (\lambda u)^4 + \Or(u^8)
    \label{LieEstimate1}
\end{equation}
when $\mathsf{u}=0$. The quadratic growth estimate here is typically
quite good even near the coordinate singularities. There is little
qualitative change in the nature of this expression if $\mathsf{u}
\neq 0$. More interestingly, the gravitational current
\eqref{GravCurrent} associated with $\psi^a_{\mathsf{xy}}$ always
vanishes. This suggests that an observer would not be compelled to
ascribe any $xy$ component of angular momentum to gravitational
waves with the given form. $H^{a}{}_{[\mathsf{u}}
\sigma_{\mathsf{v}]}$ has similar properties. It satisfies an
equation almost identical to \eqref{LieEstimate1}, and the
gravitational current generated by it always vanishes.

More interesting are the remaining two Killing-type Jacobi fields
$\psi_{\mathsf{xu}}^a = 2 H^{a}{}_{[\mathsf{x}}
\sigma_{\mathsf{u}]}$ and $\psi_{\mathsf{yu}}^a = 2
H^{a}{}_{[\mathsf{y}} \sigma_{\mathsf{u}]}$. Specializing again to
the case $\mathsf{u}=0$,
\begin{eqnarray}
  \fl \quad |\mathcal{L}_{\psi_{\mathsf{xu}}} g_{ab}|^2 \simeq |\mathcal{L}_{\psi_{\mathsf{yu}}} g_{ab}|^2 \simeq \frac{2}{3}
  \lambda^4 u^2 \Big(
  [ (x - \mathsf{x})^2+(y - \mathsf{y})^2] + \frac{2}{3} (v-\mathsf{v}) u \Big) + \Or(u^4) .
\end{eqnarray}
Unlike the expansion in \eqref{LieEstimate1}, this approximation
fails long before $|\lambda u| \rightarrow \pi/2$. In general, the
two magnitudes on the left-hand side have distinct behaviors that
strongly depend on $\mathsf{u}$. Oscillations generically arise as
$u$ is varied, for example. See Fig. 1.

\begin{figure}
    \begin{center}

    \includegraphics[width=0.8\textwidth]{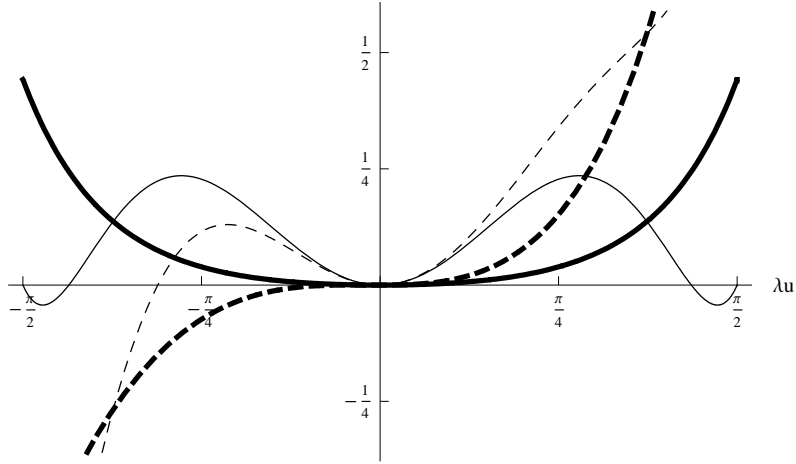}

    \caption{Plots of $|\mathcal{L}_{\psi_{\mathsf{xu}}} g_{ab}|^2$ for
    a gravitational plane wave described by \eqref{ppGeneral} and
    \eqref{GravWave}. The origin is assumed to be at $\mathsf{u}=0$.
    Both solid curves assume that $v-\mathsf{v} = 0$. The dashed ones
    use $\lambda(v-\mathsf{v})=1/2$ instead. Both thicker curves set
    $x-\mathsf{x}=0$ and $\lambda(y-\mathsf{y})=1/4$. The thinner ones use
    $\lambda(x-\mathsf{x})=1$ and $y-\mathsf{y}=0$. Plots for
    $|\mathcal{L}_{\psi_{\mathsf{yu}}} g_{ab}|^2$ look very similar
    unless $\mathsf{u} \neq 0$.}
    \end{center}
\end{figure}

There are gravitational currents associated with both
$\psi^{a}_{\mathsf{xu}}$ and $\psi^{a}_{\mathsf{yu}}$. Applying
\eqref{GravCurrent} and expanding near $\mathsf{u}=0$,
\begin{equation}
  j^a_{\psi_{\mathsf{xu}}} = \frac{\lambda^6}{180 \pi} u^4
  \left[ u \frac{\partial}{\partial x} + (x - \mathsf{x}) \frac{\partial}{\partial v} \right] + \Or (u^6) .
\end{equation}
Similarly,
\begin{equation}
  j^a_{\psi_{\mathsf{yu}}} = - \frac{\lambda^6}{180 \pi} u^4
  \left[ u \frac{\partial}{\partial y} + (y - \mathsf{y}) \frac{\partial}{\partial v} \right] + \Or(u^6).
\end{equation}
These expansions are qualitatively accurate throughout the region of
interest. It is now clear from \eqref{PStar2} and
\eqref{KomarMoment} that the only non-vanishing gravitational
momenta (associated with Killing-type GACs) are
\begin{equation}
  S^*_{\mathsf{x u}} = \int_\Sigma j^a_{\psi_{xu}} \rmd S_a ; \qquad  S^*_{\mathsf{y u}}
  = \int_\Sigma j^a_{\psi_{yu}} \rmd S_a .
\end{equation}
The rates at which these quantities change depends on the relevant
fluxes through $\partial \Sigma$. Regardless, the magnitude of the
angular momentum tensor always vanishes. Intuitively, these
statements might be taken to mean that the gravitational wave has a
``mass dipole moment'' equal to its one non-vanishing component of
ordinary angular momentum.

The results here could be straightforwardly extended to much more
general \textit{pp}-wave (and other) spacetimes. The most
interesting point is perhaps the calculation of explicit
gravitational currents in a vacuum spacetime. In these cases, the
general results obtained in Sect. \ref{Sect:ExtendedMatter} only
state that $j^a$ will decrease no slower than $X^3$ as $X
\rightarrow 0$. The example here scales like $X^5$. Although this
conclusion would probably not be preserved in more complicated
spacetimes, it shows that momenta not arising from stress-energy
tensors can sometimes be ignored in remarkably large regions.

\section{Conclusions}

Two different notions of approximate affine collineations have been
introduced. One has the physical interpretation of capturing
symmetry principles in a normal neighborhood of a point, while the
other is adapted to the measurements of a particular observer. Flows
generated by both of these objects leave $\sigma^\sfa(x,\gamma)$
invariant. This has the simple interpretation that Jacobi fields
preserve Riemann normal coordinates. GACs do the same for the
spatial components of Fermi normal coordinate systems. These objects
always exist, and each forms a 20-dimensional group. Individual
elements may be interpreted using the values of the field and its
first derivatives at the appropriate base point. The only caveat to
this is that a GAC which might initially appear to be purely
translational could slowly acquire some rotational and boost-type
components. This mixing is essential in order to ensure that the
fields nearly satisfy \eqref{AffineDef} near the observer's
worldline.

The relevance of these definitions ultimately lies in their
applications. The approximate symmetries introduced here have been
used to write down analogs of the typical conservation laws applying
to geodesics in spacetimes admitting affine collineations. Some of
the resulting parameters are exact constants of motion, while others
are only expected to vary slowly near the preferred point or
observer. Regardless, they may be used to classify and derive
geodesics in certain regions. Similar results have also been
discussed in connection with extended matter distributions. This led
to natural notions for the linear and angular momenta of a spacetime
volume as viewed in a particular frame. There seems to be some
disagreement with other quasilocal notions of gravitational momenta,
so it is not clear how the definition here should be interpreted. It
is unknown if it has any positivity or related characteristics.

Concepts discussed here might also be applied to simplify
perturbation theory off of some background geometry possessing an
exact affine collineation. It could be useful, for example, to
uniquely construct GACs with respect to a center-of-mass worldline
that coincides with exact timelike or axial Killing fields in the
unperturbed geometry. Center-of-mass trajectories might also be
estimated using notions of approximate stationarity. More
concretely, an analysis of the quantities $\ItP_\xi$ defined in
\eqref{PDef} can be shown to provide significant insights into the
effects of self-forces and self-torques on isolated bodies. Details
are presented elsewhere \cite{HarteFuture}.

\ack

I am grateful for many helpful discussions and comments from Robert
Wald and Samuel Gralla. This work was supported by NSF grant
PHY04-56619 to the University of Chicago.

\appendix

\section*{Appendix: Properties of Jacobi fields}

\setcounter{section}{1}

The Jacobi fields and generalized affine collineations satisfy a
number of useful identities that both simplify various calculations
and further motivate their identification as generators of
approximate symmetries. The most basic of these results is probably
\eqref{JacobiAffine} or its analog \eqref{GACAffine}. Symmetries
defined with respect to a given point are exactly affine at that
point. Symmetries defined with respect to a given worldline are
exactly affine on that worldline. Nothing is said about the behavior
of these fields away from these regions. It would be much more
interesting if constraints could be placed on $\nabla_a \LieS
g_{bc}$ throughout the volume $N$ where it is defined. This is
indeed possible. Jacobi fields generate diffeomorphisms that exactly
preserve certain geometric objects in all locally well-behaved
spacetimes. This fact can be used to derive some properties of
$\LieS g_{ab}$ and its derivatives even when $x \neq \gamma$.
Somewhat weaker comments can also be made regarding each GAC
$\xi^a$. Beyond this, approximations for these vector fields can be
generically obtained near their origins. A few results of this type
have been derived before for Killing-type Jacobi fields \cite{Dix74,
Dix79}. These will be generalized here while also introducing a
number of apparently new results.

The fundamental starting point is the observation that
$\sigma^\sfa(x,\gamma)$ must remain unchanged if both of its
arguments are perturbed with a Jacobi field defined about $\gamma$
\cite{Dix79}. More concisely,
\begin{equation}
  \LieS \sigma^\sfa(x,\gamma) = \psi^a \sigma^{\sfa}{}_a + A^\sfb \sigma^{\sfa}{}_\sfb - \sigma^\sfb B_{\sfb}{}^{\sfa} =
  0. \label{LieSig1}
\end{equation}
This means that Riemann normal coordinate grids are left invariant
under infinitesimal transformations generated by $\psi^a$. It can be
verified using (\ref{HDef}) and (\ref{InitialData}). The solutions
to (\ref{LieSig1}) are seen to all have the form
(\ref{JacobiFirstDer}) derived for the Jacobi fields. The
undifferentiated world function is left unchanged only under the
action of Killing-type Jacobi fields. More generally,
\begin{equation}
  \LieS \sigma(x,\gamma) = \sigma^\sfa \sigma^\sfb B_{(\sfa \sfb)} .
  \label{LieSig2}
\end{equation}
In the special case of a generalized homothetic vector field
satisfying (\ref{HomB}), this reduces to the particularly simple
form
\begin{equation}
  \Lie_{\psi_{\mathrm{H}}} \sigma = 2 c \sigma .
\end{equation}

Returning to the general case, several additional identities may be
derived from (\ref{LieSig1}) and (\ref{LieSig2}) by differentiation.
It is clear that a single covariant derivative with respect to $x$
will commute with the Lie derivatives in both of these equations.
Using this fact shows that
\begin{equation}
  \LieS \sigma^{\sfa}{}_{a} = \LieS H^{a}{}_{\sfa} = 0,
  \label{LieSig3}
\end{equation}
and
\begin{equation}
  \LieS \sigma_a = 2 \sigma^{\sfa} \sigma^{\sfb}{}_{a} B_{(\sfa \sfb)} .
  \label{LieSig4}
\end{equation}
It then follows from (\ref{SigIdent1}) and (\ref{LieSig3}) that
\begin{equation}
  \LieS \sigma^a = 0 . \label{LieSig5}
\end{equation}
Comparing this to what would be expected from (\ref{LieSig4})
produces the very useful identity
\begin{equation}
  \sigma^a \LieS g_{ab} = 2 \sigma^\sfa \sigma^{\sfb}{}_{b}
  B_{(\sfa \sfb)} . \label{LieSig6}
\end{equation}
This result is central to the interpretation of Jacobi fields as
generalized symmetry generators. It applies most clearly to
Killing-type vectors $\psi^a_{\mathrm{K}}$. The right-hand side then
vanishes, providing a strong restriction on how much these fields
can fail to satisfy Killing's equation away from $\gamma$.

More generally, there should exist an analog of (\ref{LieSig6}) that
constrains $\nabla_a \LieS g_{bc}$ for all Jacobi fields.
Expressions of this form are easily obtained by directly
differentiating (\ref{LieSig6}). Applying identities like
(\ref{SigIdent1}) to the result then shows that
\begin{equation}
   \sigma^a \sigma^b [ 2 \nabla_{(a} \LieS g_{b)c} - \nabla_c \LieS
   g_{ab} ] = 0
   \label{LieSig7}
\end{equation}
and
\begin{equation}
  \sigma^a \sigma^c \nabla_a \LieS g_{bc} = 2 \sigma^\sfa (\sigma^{\sfb}{}_{b}
  - \sigma^{\sfb}{}_{c} \sigma^{c}{}_{b} ) B_{(\sfa \sfb)} .
  \label{LieSig8}
\end{equation}
The first of these equations ensures that that there is always some
sense in which every Jacobi field is approximately affine. It is
actually identical to \eqref{Jacobi}. One trivial consequence is
that
\begin{equation}
  \sigma^a \sigma^b \sigma^c \nabla_a \LieS g_{bc} = 0 .
  \label{LieSig9}
\end{equation}
It means that the ``purely radial'' component of \eqref{AffineDef}
holds everywhere that $\psi^a$ is defined. Similar statements can
also be made using \eqref{LieSig8}. The right-hand side of that
equation vanishes for Killing or homothetic-type Jacobi fields,
although interesting remarks can be made even in the general case.
Temporarily consider points that are spacelike-separated from
$\gamma$, so $\sigma > 0$. It is then clear that in normal
coordinates, the appropriate components of $\nabla_{a} \LieS g_{bc}$
fall off roughly like $1/\sqrt{\sigma}$ far away from the origin.
These quantities also have a simple scaling behavior near $\gamma$.
Given that all third derivatives of the world function vanish when
$x \rightarrow \gamma$, the term
\begin{equation}
  \sigma^{\sfb}{}_{b} - \sigma^{\sfb}{}_{c} \sigma^{c}{}_{b} =
  \sigma^a \sigma^{\sfb}{}_{ab}
\end{equation}
in (\ref{LieSig8}) must go to zero at least as fast as $\sigma$ in
this limit. It then follows that the portions of $\nabla_{a} \LieS
g_{bc}$ constrained in that equation should grow no faster than
$\sqrt{\sigma}$ for $x$ very near $\gamma$.

A more explicit connection can actually be made between $\nabla_a
\LieS g_{bc}$ and derivatives of the world function. Several remarks
have already been made about the interpretation of various results
here in terms of normal coordinate systems. Since (\ref{AffineDef})
is essentially the condition that leaves the Levi-Civita connection
invariant, one might expect an appropriate result to arise from
studying the Christoffel symbols in normal coordinates. These are
related to the bitensor $\sigma^{\sfa}{}_{ab}$. Lie derivatives of
this quantity can be computed by directly differentiating
(\ref{LieSig3}). Simplifying the result requires the relation
\begin{equation}
  \nabla_b \nabla_a \psi_c = R_{cabd} \psi^d + \frac{1}{2} [ 2 \nabla_{(a} \LieS
  g_{b)c} - \nabla_c \LieS g_{ab} ]
  .
    \label{Del2General}
\end{equation}
This is obtained by repeatedly applying Ricci's identity, and
actually applies for any vector field. It is a generalization of
(\ref{Del2Affine}) and the well-known expression for second
derivatives of a Killing field. Combining (\ref{HDef}),
(\ref{LieSig3}), and (\ref{Del2General}) shows that
\begin{equation}
  \nabla_a \LieS g_{bc} = 2 H_{(c|\sfa|} \LieS \sigma^{\sfa}{}_{b)a}
  . \label{DelLieGStart}
\end{equation}
This is essentially the expected result.

There is at least one more quantity derived from the world function
that has a simple interpretation in normal coordinates. That is the
van Vleck determinant
\begin{equation}
  \Delta(x, \gamma) = \frac{ \det( - \sigma_{\sfa a} ) }{\sqrt{-g(x)}
  \sqrt{-g(\gamma)}} . \label{DeltaDef}
\end{equation}
This two-point scalar arises as part of the volume element in these
charts. It also appears in connection with the focusing of geodesic
congruences and in Green functions associated with common wave
equations \cite{PoissonRev}. It is clear from the definition
(\ref{DeltaDef}) that
\begin{equation}
  \nabla_b \ln \Delta = - H^{a}{}_{\sfa} \sigma^{\sfa}{}_{ab} .
\end{equation}
Derivatives with respect to $\gamma$ are identical if the index $b$
in this equation is changed to a $\sfb$. Simplifying with
(\ref{LieSig3}) then shows that
\begin{equation}
  \LieS \ln \Delta = \nabla_\sfa \psi^\sfa - \nabla_a \psi^a =
  B^{\sfa}{}_{\sfa} - \frac{1}{2} g^{ab} \LieS g_{ab} .
\end{equation}

The results obtained so far apply everywhere that the generalized
symmetries are defined. It is sometimes more useful to derive
explicit approximations for the Jacobi fields. These can be found in
various ways. Perhaps the most obvious procedure is to evaluate
\eqref{JacobiFirstDer} in a normal coordinate system with origin
$\gamma$. The second term involving $B^{\sfb \sfa}$ then reduces to
a linear combination of the coordinate functions. This is exact. The
factor of $\sigma^{\sfa}{}_{\sfb}$ appearing in the other term does
not simplify in these coordinates. It reduces to the identity at the
origin, although its behavior further away must be computed in some
other way. For $X^\sfa$ very small, known Taylor expansions for
$\sigma^{\sfa}{}_{\sfb}$ could be used \cite{PoissonRev}. There are
also integral approximations neglecting all terms nonlinear in the
curvature \cite{Synge, deFelice}. Either of these methods can be
used to find explicit expressions for $\psi^a$ or $\xi^a$ in
appropriate situations, although it is often more interesting to
directly compute Lie derivatives of the metric with respect to these
vector fields. The viewpoint here will be to derive a Taylor series
for $\LieS g_{ab}$ centered at $\gamma$. This expansion will be in
powers of the radial vector $-\sigma^\sfa$. The first two terms are
clear from \eqref{InitialData} and \eqref{GACAffine}. Going to
higher orders requires knowledge of $[\nabla_a \nabla_b \LieS
g_{cd}]$, $[\nabla_a \nabla_b \nabla_c \LieS g_{df}]$, and so on.
The bracket notation here was defined in \eqref{CoincDef}. In any
case, limits like these may be computed from \eqref{DelLieGStart}.
Results such as \eqref{Del2General} can further be used to derive
the full third or fourth covariant derivatives of $\psi^a$ in the
coincidence limit.

First consider third derivatives. From \eqref{Del2General},
$\nabla_a \nabla_b \LieS g_{cd}$ involves $[\nabla_b
H^{a}{}_{\sfa}]$ and $[\nabla_d \LieS \sigma^{\sfa}{}_{bc}]$. All
third derivatives of the world function vanish at coincidence, so
only the second term is important here. Some manipulation shows that
\begin{equation}
  [\nabla_d \LieS \sigma^{\sfa}{}_{bc} ] =  \LieS [ \sigma^{\sfa}{}_{bcd}
  ] .
\end{equation}
The quantity being differentiated on the right-hand side is known to
have the form \cite{PoissonRev, Synge, deFelice}
\begin{equation}
[\sigma^{\sfa}{}_{bcd}] = -\frac{2}{3} R^{\sfa}{}_{(bc)d} ,
\label{Sig4Coinc}
\end{equation}
so
\begin{equation}
  [\nabla_a \nabla_b \LieS g_{cd} ] =  \frac{2}{3} \big( R_{\sfa \sfb(\sfc}{}^{\sff}
  \LieS g_{\sfd) \sff} + g_{\sff ( \sfc} \LieS R_{\sfd) \sfa \sfb}{}^{\sff} \big) .
  \label{Del2Lieg}
\end{equation}
Lie derivatives here have their usual expansions in terms of
$A_\sfa$, $B_{\sfa \sfb}$, $R_{\sfa \sfb \sfc}{}^{\sfd}$, and
$\nabla_\sff R_{\sfa \sfb \sfc}{}^{\sfd}$. The first term clearly
vanishes for Killing-type Jacobi fields. The other depends on how
close $\psi^a$ comes to being a curvature collineation (satisfying
$\Lie_{Y_{\mathrm{C}}} R_{abc}{}^{d} =0$) at $\gamma$.

It is not too difficult to compute one more derivative of $\LieS
g_{ab}$. This is again derived from \eqref{DelLieGStart}. Noting
that
\begin{equation}
  [\LieS \sigma^{\sfa}{}_{b c} ] = \LieS [ \sigma^{\sfa}{}_{b c} ] =
  0,
\end{equation}
one finds
\begin{equation}
  [\nabla_a \nabla_b \nabla_c \LieS g_{df} ] = 2 [\nabla_a \nabla_b
  \LieS \sigma^{\mathsf{h}}{}_{c(d} ] g_{\sff) \mathsf{h}} .
\end{equation}
The two covariant derivatives here may be commuted inside the Lie
derivative. A number of remainder terms arise, although these all
vanish on account of \eqref{Del2General} and \eqref{Sig4Coinc}. This
leaves an expression dependent only on $[\sigma^{\sfa}{}_{b c d
f}]$. As noted in \cite{Synge},
\begin{equation}
  [\sigma_{abcdf}] = \nabla_{(\sff} R_{\sfd)(\sfa \sfb) \sfc} + \frac{1}{2} \nabla_\sfc
  R_{\sfd ( \sfa \sfb) \sff} .
\end{equation}
A straightforward application of Synge's rule together with
\eqref{Sig4Coinc} shows that
\begin{equation}
  [\sigma^{\sfa}{}_{bcdf}] = \frac{1}{6} \nabla^\sfa R_{\sfb ( \sfd
  \sff) \sfc} - \nabla_{(\sff} R_{\sfd) (\sfb \sfc)}{}^{\sfa} .
\end{equation}
Combining all of these results,
\begin{eqnarray}
  [\nabla_a \nabla_b \nabla_c \LieS g_{df} ] =& \frac{1}{3}
  g_{ \mathsf{h} (\sfd} \LieS \nabla^{\mathsf{h}} R_{\sff) (\sfa
  \sfb) \sfc} - g_{\mathsf{h} \sfd} \LieS \nabla_{(\sfa} R_{\sfb)
  (\sfc \sff)}{}^{\mathsf{h}}
  \nonumber \\
  ~ & - g_{\mathsf{h} \sff} \LieS \nabla_{(\sfa} R_{\sfb)
  (\sfc \sfd)}{}^{\mathsf{h}} .
  \label{Del3Lieg}
\end{eqnarray}
Taylor expansions derived from these limits are given by
\eqref{LieGExpandJacobi}, \eqref{LieGGradJacobi},
\eqref{LieGExpandGAC}, and \eqref{LieGGradGAC}.

\end{document}